\documentclass[a4paper,fleqn,review]{cas-dc}

\usepackage[numbers,sort&compress]{natbib}
\usepackage{mathtools}

\DeclarePairedDelimiterX\braket[2]{\langle}{\rangle}{#1 \delimsize\vert #2}
\DeclarePairedDelimiterX\expval[3]{\langle}{\rangle}{#1 \delimsize\vert #2 \delimsize\vert #3}

\def\ca{c_\alpha}
\def\sa{s_\alpha}
\def\cb{c_\beta}
\def\sb{s_\beta}
\def\cg{c_\gamma}
\def\sg{s_\gamma}
\def\s{\sigma}

\def\al{$^{27}$Al}
\def\vv{$^{51}$V}
\def\mn{$^{55}$Mn}
\def\pb{$^{207}$Pb}
\def\cd{$^{113}$Cd}
\def\etal{\emph{et al.}}
\def\Circle{\raisebox{-0.33em}{\scalebox{2.5}{$\bullet$}}}

\usepackage{euscript}

\renewcommand{\H}{\hat{\EuScript{H}}}
\newcommand{\I}{\hat{I}}

\begin{document}
\let\WriteBookmarks\relax
\def\floatpagepagefraction{1}
\def\textpagefraction{.001}
\shorttitle{Single-Crystal NMR}
\shortauthors{T. Vosegaard}

\title [mode = title]{Single-Crystal NMR Spectroscopy}                      
\tnotemark[1]

\tnotetext[1]{Edited by Geoffrey Bodenhausen and Dominique Massiot}

%\tnotetext[2]{The second title footnote which is a longer text matter
%   to fill through the whole text width and overflow into
%   another line in the footnotes area of the first page.}

\author[1]{Thomas Vosegaard}[orcid=0000-0001-5414-4550]
\cormark[1]
\ead{tv@chem.au.dk}

\address[1]{Department of Chemistry and Interdisciplinary Nanoscience Center, Aarhus University, Gustav Wieds Vej 14, DK-8000 Aarhus C, Denmark}

\cortext[cor1]{Corresponding author}

\begin{abstract}
Single-crystal (SC) NMR spectroscopy is a solid-state NMR method that has been used since the early days of NMR to study the magnitude and orientation of tensorial nuclear spin interactions in solids. This review first presents the field of SC NMR instrumentation, then provides a survey of software for analysis of SC NMR data, and finally it highlights selected applications of SC NMR in various fields of research. The aim of the last part is not to provide a complete review of all SC NMR literature but to provide examples that demonstrate interesting applications of SC NMR.
\end{abstract}

%\includegraphics{Graphical_abstract-eps-converted-to.pdf}
%\item Instrumentation for single-crystal NMR spectroscopy is reviewed
%\item The theoretical basis for analysis of single-crystal NMR is summarized
%\item Software tools to analyse single-crystal NMR data are reviewed
%\item Selected applications of single-crystal NMR in materials science are presented
%\item Applications of single-crystal NMR to study various physical properties are presented


\maketitle

\tableofcontents
%%%%%%%%%%%%%%%%%%%%%%%%%%%%%%%%%%%%%%%%%
% INTRODUCTION 
%%%%%%%%%%%%%%%%%%%%%%%%%%%%%%%%%%%%%%%%%
\section{Introduction}\label{sec:intro}

NMR has been used to investigate materials in the solid phase for more than seven decades \cite{Purcell_Torrey_Pound_1946}. In the early days of NMR, many studies were carried out on single-crystal samples, as such samples in general simplify matters significantly compared to powders. One of the first NMR studies of powdered samples was the famous observation of the so-called Pake doublet revealing the dipole-dipole coupling between the two hydrogen atoms in water in gypsum \cite{Pake_1948}. Although Pake studied a gypsum powder, he also reported the peak splitting for different orientations of a single crystal. Following significant developments probe technology for magic-angle spinning NMR probes and of computer software for numerical simulations of solids, the number of SC NMR studies is declining as solid-state NMR investigations of powders become easier to perform. However, researchers still turn to SC NMR when precise determination of nuclear spin interaction parameters is required or when the orientation of the tensorial interaction needs be determined relative to the molecular coordinate system.

%%%%%%%%%%%%%%%%%%%%%%%%%%%%%%%%%%%%%%%%%
% PROBES 
%%%%%%%%%%%%%%%%%%%%%%%%%%%%%%%%%%%%%%%%%
\section{Single-Crystal NMR Probes}\label{sec:probes}

In the early days of NMR, hardware developments went hand in hand with experiments and applications. Many different groups were building their own NMR probes, but this trend is gradually changing as commercial equipment has become better. In addition, as the magnetic fields become higher and higher, the radio-frequency (RF) parts of the probes becomes more challenging to design. For SC NMR, the commercial options are very limited, so there is still a need for designing one's own hardware for this discipline.

Many different hardware designs have been presented for SC NMR experiments over the years. The major challenges are $(i)$ to perform the stepwise rotation of the crystal in a precise manner and $(ii)$ be able to relate the orientation of the crystal inside the magnet to the orientation of the unit cell axes of the crystal -- which is normally done by X-ray diffraction (XRD) or optical face indexing on the same crystal in a separate measurement. This chapter gives an overview of the most common SC NMR probe designs but also reviews some of the more exotic hardware designs and add-ons to extend the capability of existing NMR probes for SC NMR studies.

% Dedicated probes
\subsection{Dedicated Single-Crystal NMR Probes}

\begin{figure}[pos=t]
	\centering
	\includegraphics[width=0.7\linewidth]{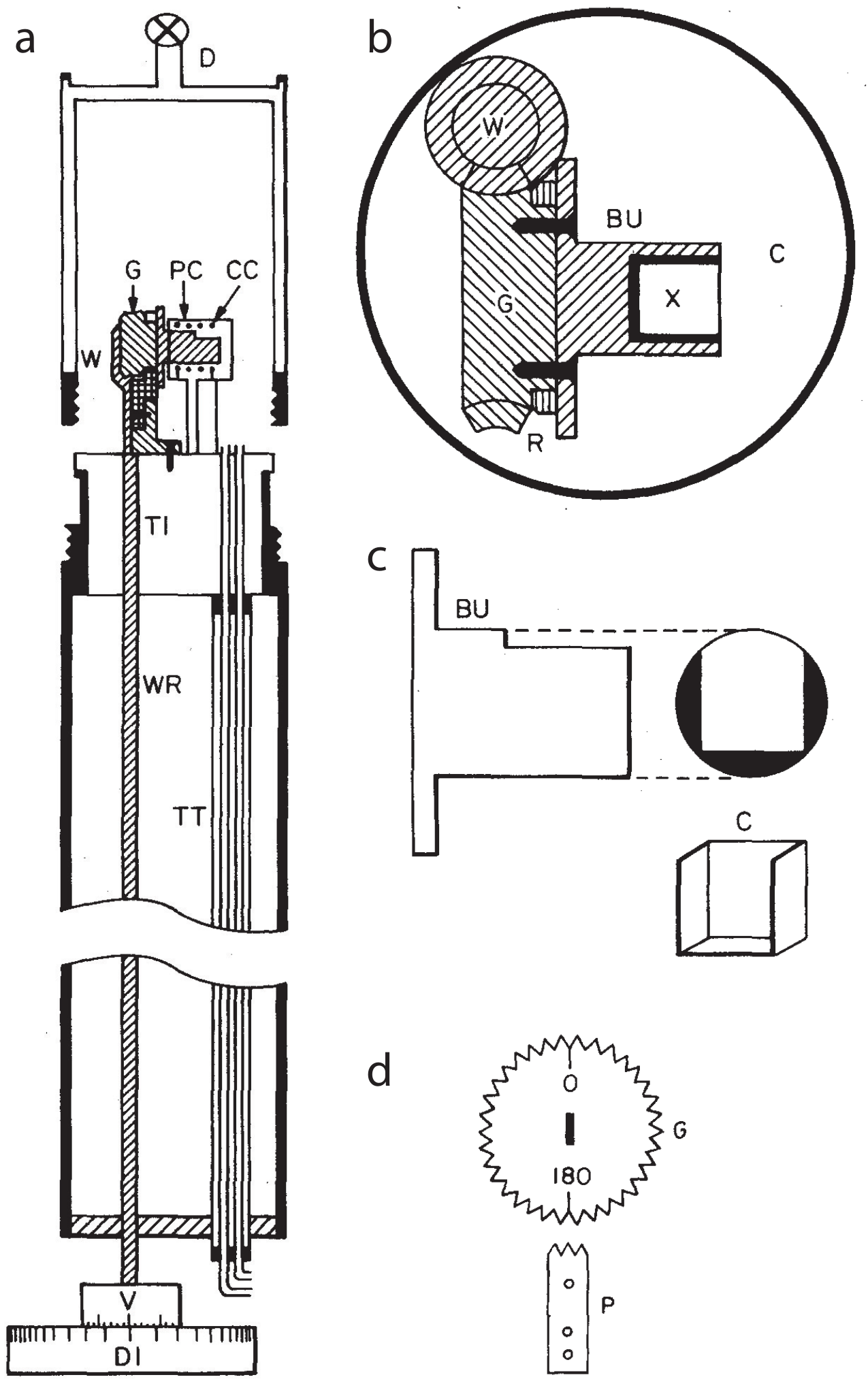}
	\caption{Common SC NMR probe design. (a) Side view and (b) top view of the probe of Waugh and co-workers \cite{Pausak_Pines_Waugh_1973, Pausak_Tegenfeldt_Waugh_1974} showing the worm gear, coil, crystal holder etc. For a full list of the labelled features, see ref.~\citealp{Pausak_Tegenfeldt_Waugh_1974}. (c) Expansion of the crystal holder and cubic crystal container. (d) Mechanism to lock the crystal orientation at a specific position in the original design \cite{Pausak_Pines_Waugh_1973}. Adapted from \cite{Pausak_Tegenfeldt_Waugh_1974} and \cite{Pausak_Pines_Waugh_1973} with permission.}
	\label{fig:pausak}
\end{figure}

The most widely used SC NMR probe designs rely on stepwise rotation of the crystal around an axis perpendicular to the magnetic field \cite{Pausak_Pines_Waugh_1973, Pausak_Tegenfeldt_Waugh_1974, Pines_Chang_Griffin_1974, Kempf_Spiess_Haeberlen_Zimmermann_1974, Veeman_1981, Honkonen_Doty_Ellis_1983, Veeman_1984, Vosegaard_Langer_Daugaard_Hald_Bildse_Jakobsen_1996}. Figure \ref{fig:pausak} shows such a probe, designed by Pausak et al. [3,4], in which the crystal is mounted inside a four-faced cube (labelled C in Fig.~\ref{fig:pausak}c) that may be mounted in three different orthogonal orientations inside the probe. The three mountings correspond to rotations of the cube around its x, y, and z axes, respectively.

\begin{figure*}[pos=t]
	\centering
		\includegraphics[width=0.7\linewidth]{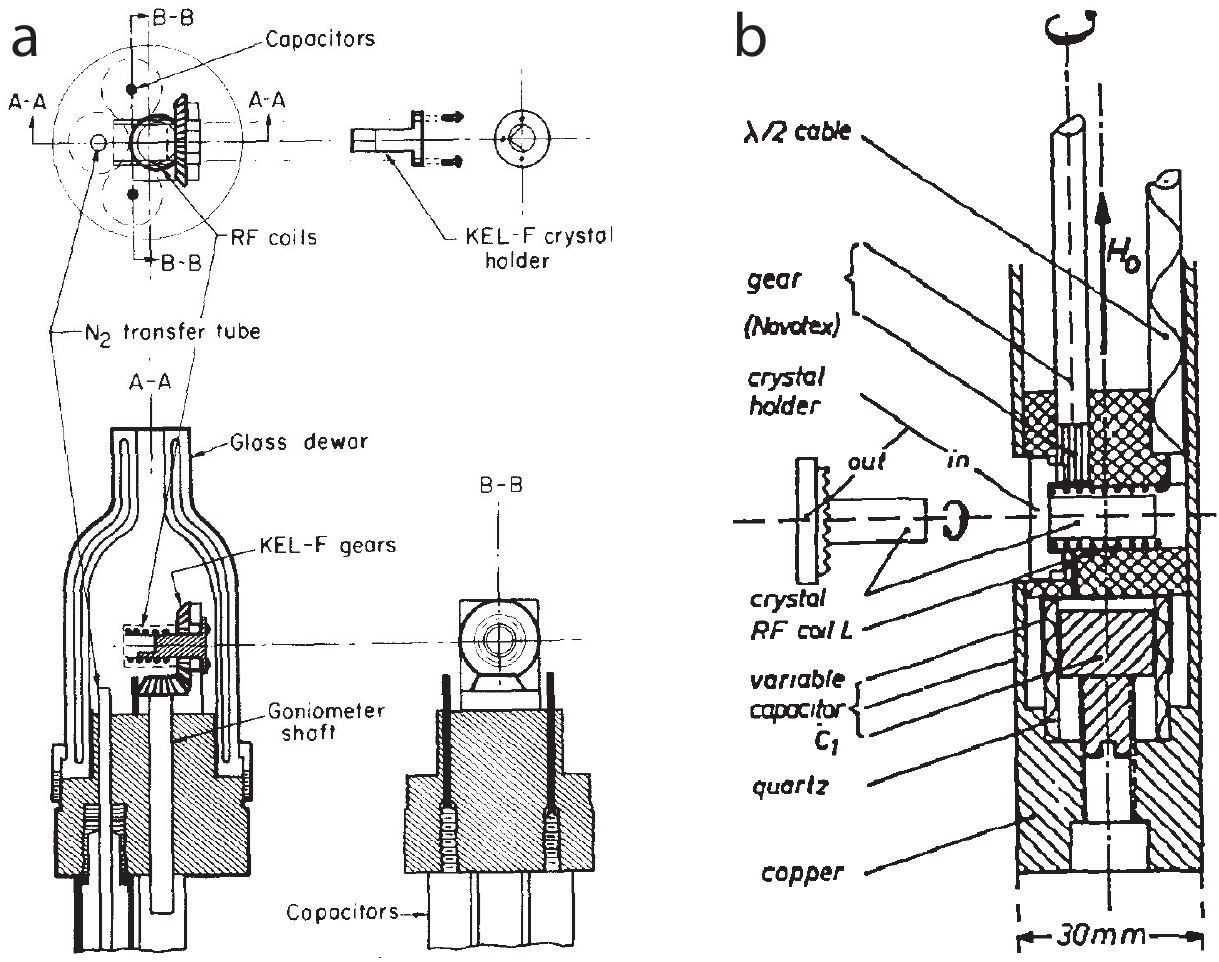}
	\caption{Two examples of SC NMR probes employing right-angle gears to achieve a horizontal rotation of the crystal (perpendicular to the vertical magnetic field) through operation from the bottom or top of the probe. Reproduced from \cite{Pines_Chang_Griffin_1974} and \cite{Kempf_Spiess_Haeberlen_Zimmermann_1974} with permission.}
	\label{fig:griffin_kempf}
\end{figure*}

\begin{figure*}[pos=t]
	\centering
		\includegraphics[width=0.6\linewidth]{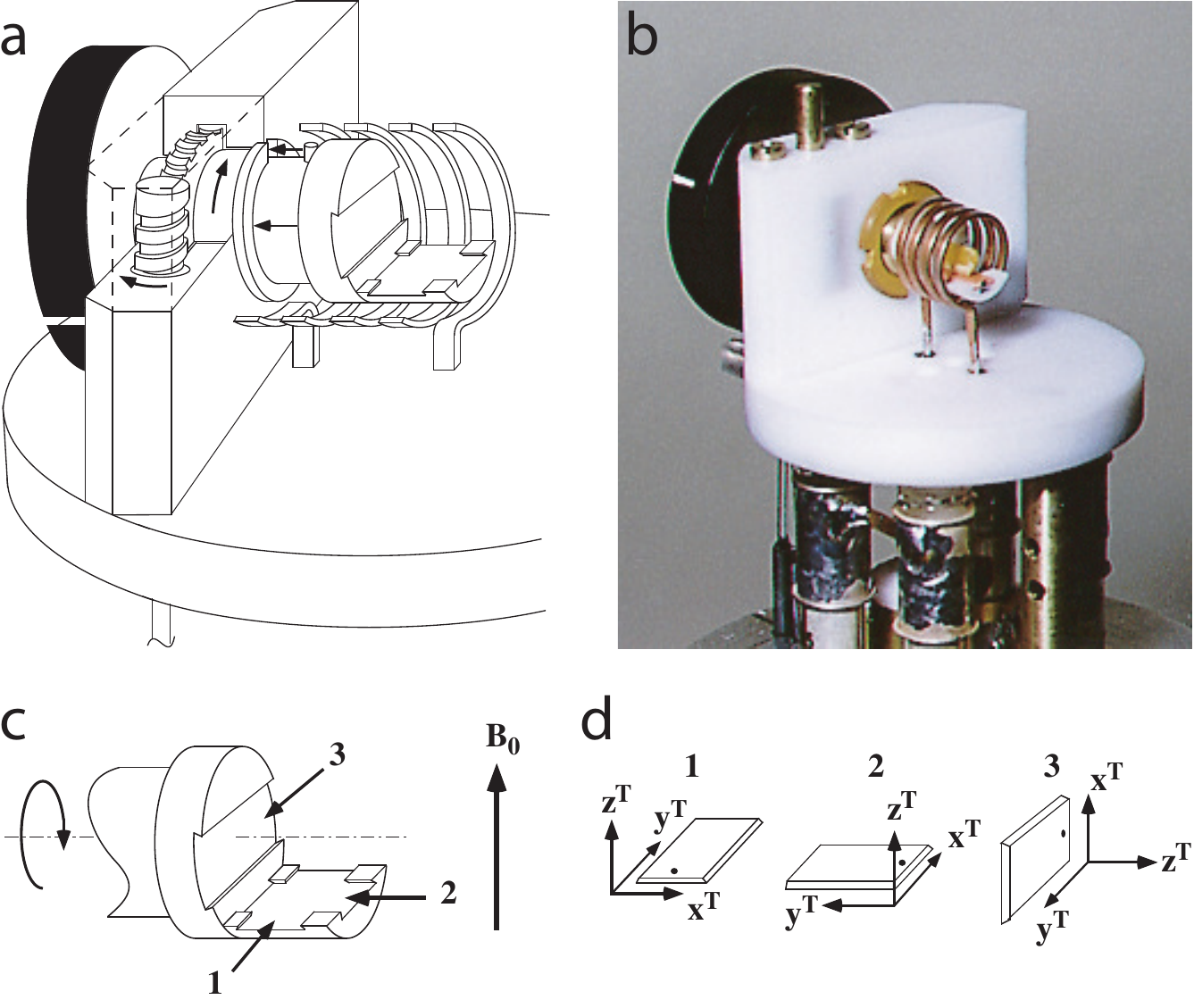}
	\caption{Our three-axis worm-gear SC goniometer design using a dovetail and mortise for precise mounting of the crystal. (a,b) Schematic and picture of the SC probe. In (b) a yellow crystal of Rb$_2$CrO$_4$ is visible inside the coil. (c,d) Illustration of the three orthogonal mountings of the tenon holding the crystal. Adapted from \cite{Vosegaard_Langer_Daugaard_Hald_Bildse_Jakobsen_1996} with permission.}
	\label{fig:vosegaard3}
\end{figure*}

Since superconducting magnets have their bore along the direction of the magnetic field, a common feature of all SC NMR probes is the need for a right-angle gear to perform the crystal rotation at an angle perpendicular to the magnetic field while operating the rotation from the bottom of the probe. There are two prevalent ways to achieve the $90^\circ$ bend: ($i$) by a set of right-angle gears or ($ii$) by a worm-gear. It should be noted that in earlier SC NMR probe designs for electromagnets, this was not necessary \cite{Veeman_1981, Veeman_1984}, as the center of the magnet could be reached from the side of the magnet.

The worm gear consists of a vertical screw on a shaft that may be turned from the bottom of the probe as illustrated with the dial in Fig.~\ref{fig:pausak}. The screw meshes with a gear connected to the crystal holder with its axis of rotation perpendicular to the axis of the probe, thus providing the perpendicular rotation axis. The advantage of a worm gear is that if the rotation is always performed in the same direction, the worm gear has no backlash, providing a very precise orientation setting.

Right-angle gears, also called bevel gears come in many different forms as illustrated by the two probes in Fig.~\ref{fig:griffin_kempf}. The probe by Griffin and co-workers \cite{Pines_Chang_Griffin_1974} used a set of conic spur gears with the teeth cut off at an angle of 45$^\circ$ to ensure a frictionless mesh, while the probe of Kempf et al.~\cite{Kempf_Spiess_Haeberlen_Zimmermann_1974} uses a slightly different design of the spur gears. While mechanical engineers often prefer gears with a helical twist of the teeth to provide a smoother mesh, this is generally not the case for SC probe designs, probably because of the small size of the gears required for SC NMR.

Precise setting of the crystal rotation angle is essential for precise measurement of the tensorial nuclear spin interaction parameters. Most conventional SC NMR probes have reported precisions of the angle setting from "a small fraction of a degree" \cite{Pausak_Tegenfeldt_Waugh_1974} and $\pm 0.07^\circ$ \cite{Calsteren_Birnbaum_Smith_1987} to $\pm 0.3^\circ$ \cite{Pausak_Pines_Waugh_1973} and $\pm 1^\circ$ \cite{Kempf_Spiess_Haeberlen_Zimmermann_1974}. While better precision is probably not required to accurately measure the NMR parameters, we have found that very high precision for the rotation and for the different orthogonal mountings of the crystal is essential to be able to correlate the rotation plots from different rotation axes. As the resonances may make excursions over hundreds of kHz per degree of rotation, any inaccuracy is clearly visible and leads to difficulties in the data analysis. Hence, we developed a SC NMR probe which eliminated any backlash by using a worm gear for the crystal rotation and a dovetail mortise for mounting the crystal \cite{Vosegaard_Langer_Daugaard_Hald_Bildse_Jakobsen_1996} as illustrated in Fig.~\ref{fig:vosegaard3}. With this probe, it has been possible to achieve precisions in the angle setting of around $\pm 0.1^\circ$. We have recently re-evaluated the precision of the old data and found that the mechanical precision of the probe design allows mounting and rotating the crystal with a precision of $\pm 0.03^\circ$ to $\pm 0.06^\circ$ \cite{Vinding_Kessler_Vosegaard_2016}.

\begin{figure}[pos=t]
	\centering
		\includegraphics[width=\linewidth]{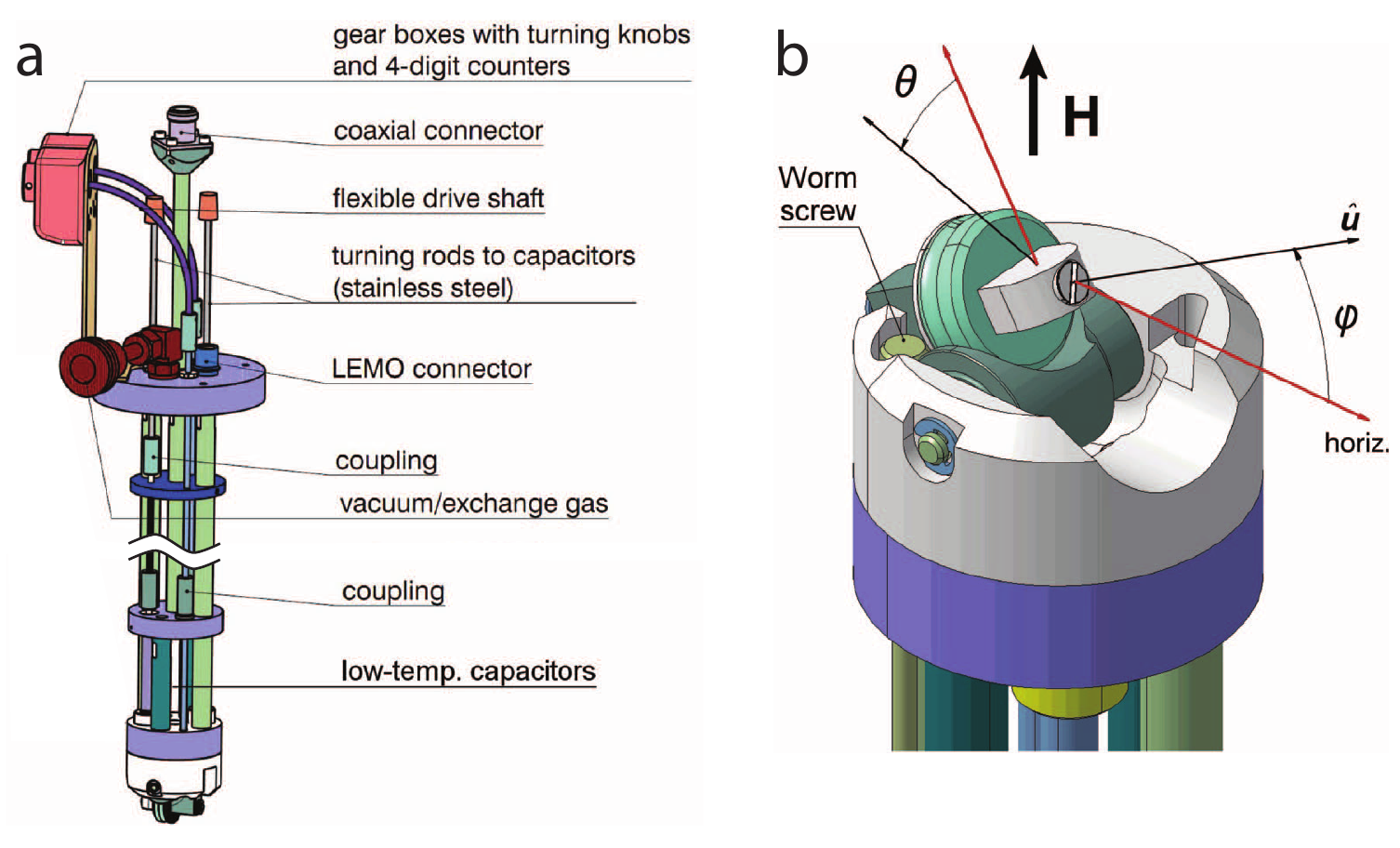}
	\caption{Low-temperature SC NMR probe. (a) Schematic of the cryostat insert with the SC goniometer at the bottom and various items for tuning, matching, sample rotation, and cooling. (b) Expansion of the goniometer head showing the axis of rotation and the rotational pitch. The probe fits into a conventional helium flow cryostat. Adapted from \cite{Shiroka_Casola_Mesot_Bachmann_Ott_2012} with permission.}
	\label{fig:low_temp}
\end{figure}

Further advances in SC NMR probe technology was done by Shiroka et al.~\cite{Shiroka_Casola_Mesot_Bachmann_Ott_2012} to enable SC NMR experiments at cryogenic temperatures. Their design, shown in Fig.~\ref{fig:low_temp}, uses a two-axis goniometer design that fits into a cryostat for helium cooling. Like most other two-axis designs, the rotation axis is not perpendicular to the magnetic field direction.

\subsection{Sensitivity Considerations}

\begin{figure*}[pos=t]
	\centering
		\includegraphics[width=0.7\linewidth]{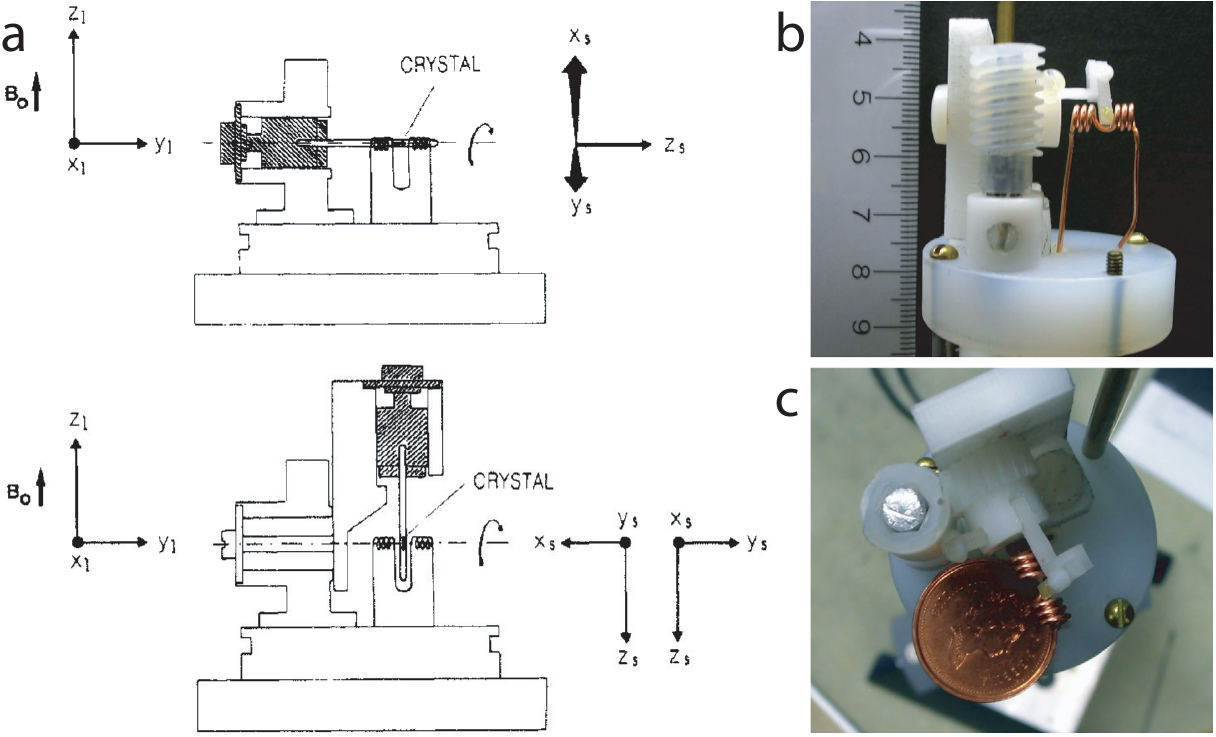}
	\caption{Split-coil SC NMR probe designs by (a) Pascher and co-workers \cite{Hauser_Radloff_Ernst_Sundell_Pascher_1988} and (b,c) Sykes and co-workers \cite{Devries_Zhao_Sykes_2007}. The photograph in (b) shows the side view of the probe, while (c) shows the top view of the probe. Reproduced from \cite{Hauser_Radloff_Ernst_Sundell_Pascher_1988} and \cite{Devries_Zhao_Sykes_2007} with permission.}
	\label{fig:split_coil}
\end{figure*}

\begin{figure}[pos=t]
	\centering
		\includegraphics[width=\linewidth]{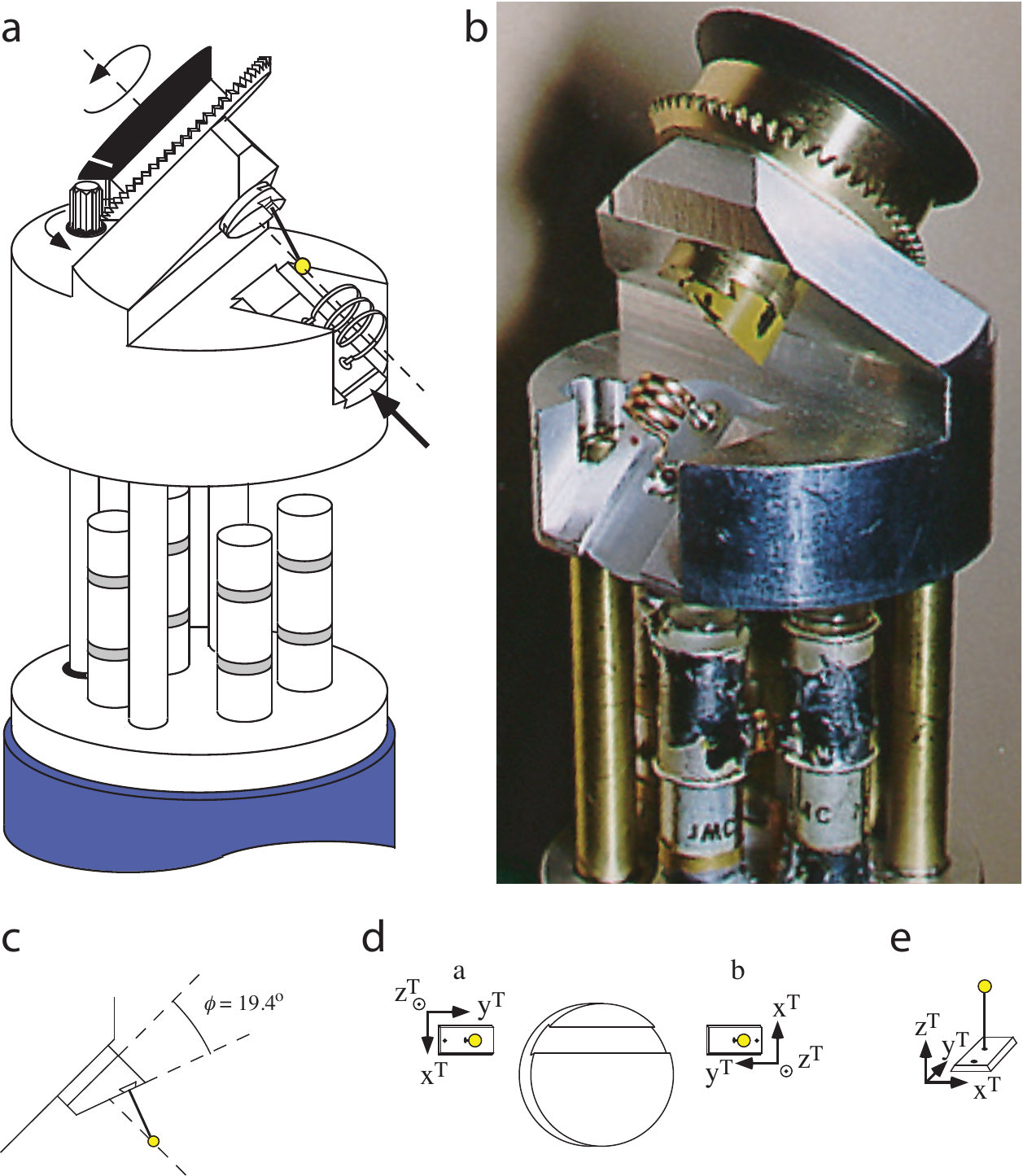}
	\caption{(a) Schematic and (b) photograph of our two-axis SC NMR probe employing a rotation axis inclined by $45^\circ$ relative to the magnetic field and with the crystal mounted on a glass stick for better filling factor of the RF coil. (c,d) Illustrations of the mounting of the crystal tenon leading to two different rotation axes marked a and b in (d). (e) Crystal (yellow sphere) mounted on a glass stick glued onto the tenon. Adapted from \cite{Vosegaard_Hald_Daugaard_Jakobsen_1999} with permission.}
	\label{fig:vosegaard2}
\end{figure}

In the SC NMR probes presented in the previous section, the crystal holder, either a open cube or a tenon, and part of the goniometer has been placed inside the RF coil. While this is the most convenient design, it provides a poor filling factor of the coil leading to low sensitivity. To improve this, Pascher and co-workers \cite{Hauser_Radloff_Ernst_Sundell_Pascher_1988} and Sykes and co-workers \cite{Devries_Zhao_Sykes_2007} proposed SC probes with a split-coil design to allow closer magnetic coupling between the crystal and the coil. These probes are shown in Fig.~\ref{fig:split_coil}. Indeed, using a split-coil design, it was possible to perform $^{31}$P SC NMR studies of sub-mm$^3$ crystals of phospholipid mimics \cite{Hauser_Radloff_Ernst_Sundell_Pascher_1988}.

We pursued a different route in moving the goniometer out of the sensitive volume of the RF coil by designing a SC NMR probe with only two axes of rotation \cite{Vosegaard_Hald_Daugaard_Jakobsen_1999}. In this design, the crystal is mounted on a glass stick which is inserted into the coil as illustrated in Fig.~\ref{fig:vosegaard2}. This ensures a very high filling factor of the coil. In order to characterise the full nuclear spin interaction tensor, the rotation angle had to differ from the standard perpendicular orientation, and an angle of $45^\circ$ was chosen in this case. Additionally, the crystal could be mounted in two different ways (marked a and b in Fig.~\ref{fig:vosegaard2}d), providing two different angular dependencies of the resonance frequencies \cite{Vosegaard_Hald_Daugaard_Jakobsen_1999}. The ability of this probe to provide good NMR sensitivity was demonstrated for $^{31}$P and $^{87}$Rb NMR studies of a tiny crystals of (HN$_4$)$_2$HPO$_4$ and RbZn$_2$(HPO$_4$)PO$_4$ \cite{Vosegaard_Daugaard_Hald_Jakobsen_2000}.

\subsection{Angle-Flipping Correlations}

\begin{figure}[pos=t]
	\centering
		\includegraphics[width=\linewidth]{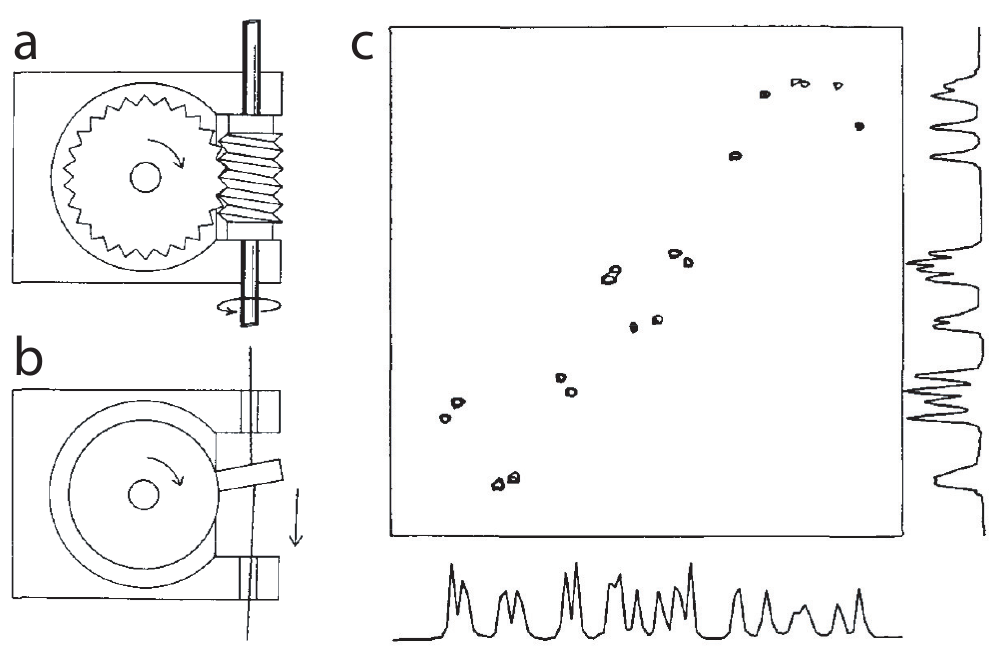}
	\caption{(a) Worm-gear SC NMR probe and (b) a modified version allowing fast flip between two different crystal orientations separated by $45^\circ$. (c) Two-dimensional $^{13}$C-$^{13}$C spectrum of 1,3,5-trimethoxybenzene correlating two different crystal orientations. Adapted  from  \cite{Carter_Alderman_Grant_1987} with permission.}
	\label{fig:carter}
\end{figure}

\begin{figure}[pos=t]
	\centering
		\includegraphics[width=\linewidth]{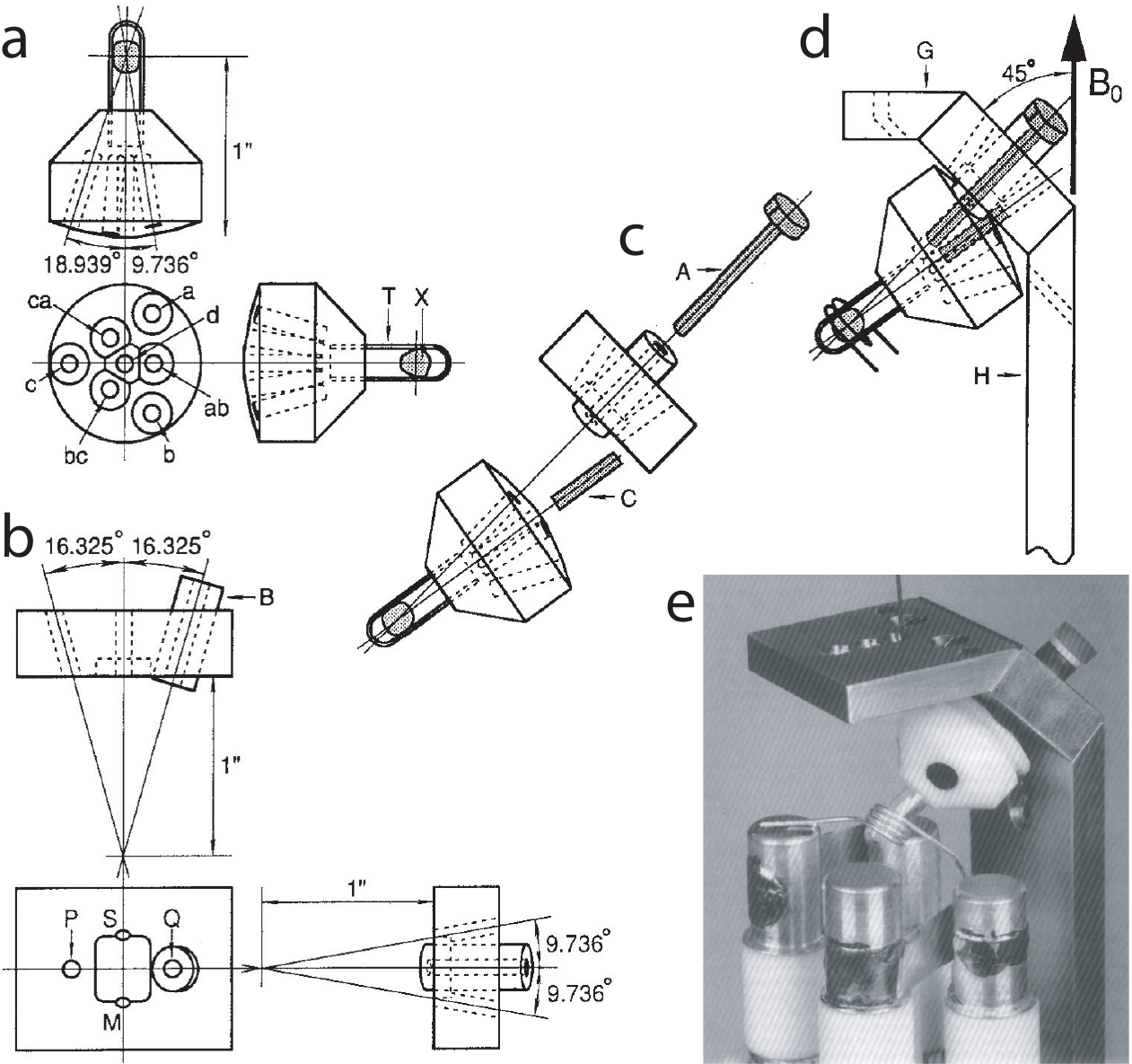}
	\caption{(a-d) Schematic of the Sherwood SC NMR probe for 2D correlation of only six different crystal orientations. (a) Armature holding the crystal, (b) aluminium plate for mounting the crystal armature. This plate is mounted at an angle of $45^\circ$ relative to the magnetic field as shown in (d). (e) Photograph of the probe. Adapted from \cite{Sherwood_Alderman_Grant_1989} with permission.}
	\label{fig:sherwood}
\end{figure}

A challenge, particularly for crystals with many different sites and/or several asymmetric units in the unit cell, is to establish an unambiguous assignment of resonances for different rotation angles in the so-called rotation plots. The simplest but time-consuming solution is to use small time steps, but even then, it may still be difficult to assign overlapping or dense regions of the spectrum and to correlate resonances from different crystal orientations. To overcome this problem, Grant and co-workers modified a conventional single-crystal NMR probe by replacing the worm gear with a flipping system \cite{Carter_Alderman_Grant_1987}. The system allowed fast flipping of the crystal between two different rotation angles as illustrated in Fig.~\ref{fig:carter}. The flipping was controlled by the spectrometer computer and fast enough to establish 2D spectra correlating the resonance frequencies at the two different orientations of the crystal (cf.~Fig.~\ref{fig:carter}c). This solved the problems of assignment of the rotation plot for each perpendicular mounting of the crystal but did not solve the problem of assigning the rotation plots of the three mountings of the crystal.

Addressing the wish to correlate all crystal orientations required for characterisation of the full chemical shift tensors, Sherwood et al.~\cite{Sherwood_Alderman_Grant_1989} presented a strategy in which only six different crystal orientations were measured. Through an ingenious goniometer design, it was possible to establish a set of 2D correlation experiments correlating all six different crystal orientations by using a variety of different mountings and flippings of the crystal. Figure \ref{fig:sherwood} shows the schematics and a photograph of the Sherwood SC NMR probe, but the reader is referred to ref.~\citealp{Sherwood_Alderman_Grant_1989} for further details. Using this goniometer design, it was possible to unambiguously assign the 12 $^{13}$C-chemical shift tensors in sucrose \cite{Sherwood_Alderman_Grant_1989, Sherwood_Alderman_Grant_1993}.

\subsection{Conventional Probes Modified for Single-Crystal NMR Studies}

\begin{figure}[pos=t]
	\centering
		\includegraphics[width=\linewidth]{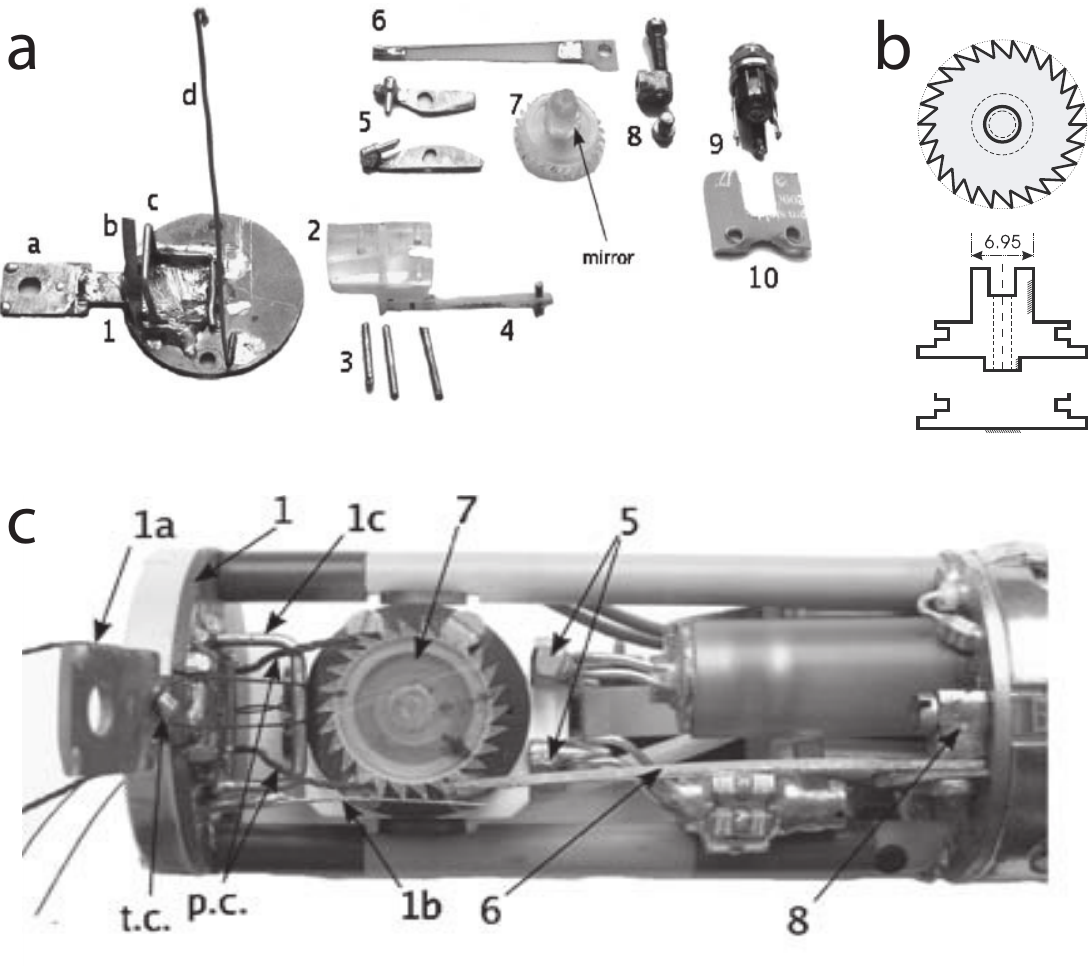}
	\caption{SC NMR kit to modify a 7 mm MAS probe \cite{Kovacs_Rohonczy_2006}. (a) Parts required for the SC NMR modification. Essential parts are the top plate (left), the ratchet pawl (6) and wheel (7) to drive the stepwise rotation of the crystal. (b) Illustration of the ratchet wheel and sample holder. (c) Picture of the rebuilt 7mm MAS probe with the stator flipped to rotate at an angle of $90^\circ$. Adapted  from \cite{Kovacs_Rohonczy_2006} with permission.}
	\label{fig:mas_probe}
\end{figure}

\begin{figure*}[pos=t]
	\centering
		\includegraphics[width=0.7\linewidth]{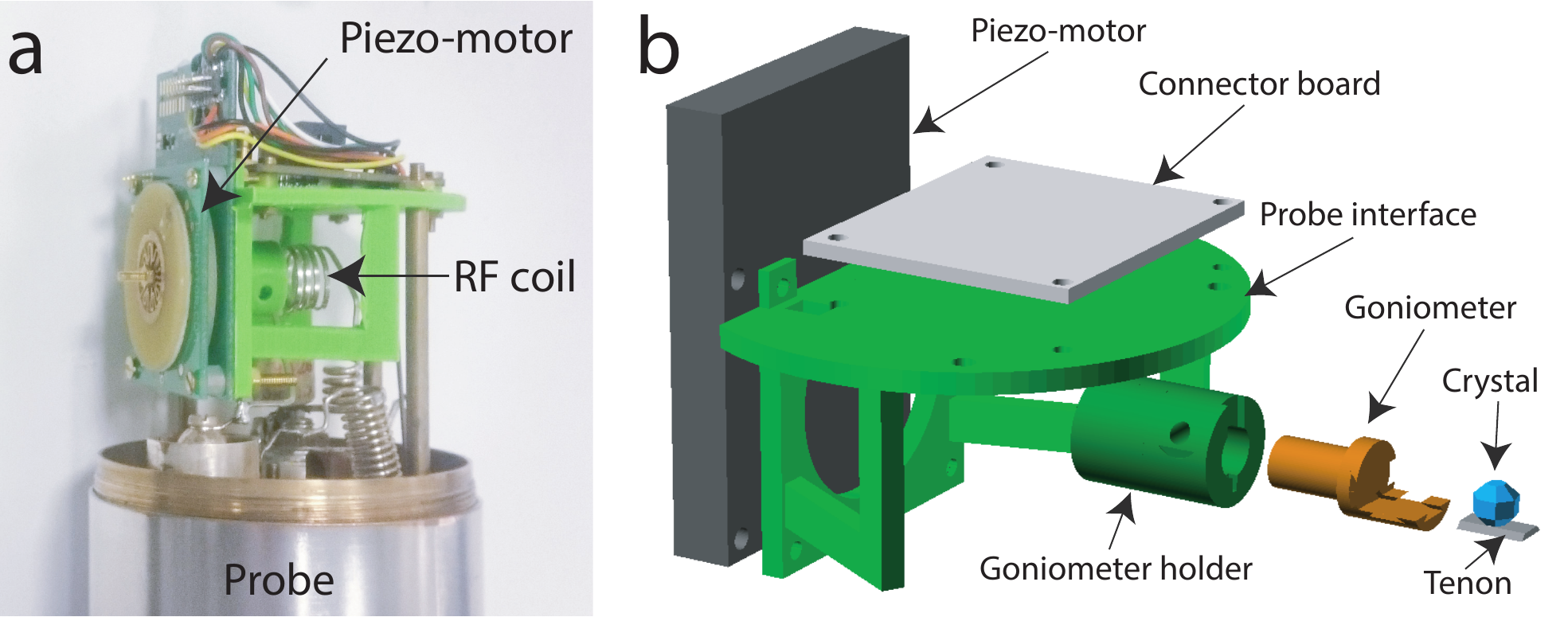}
	\caption{SC NMR kit for a standard static NMR probe \cite{Vinding_Kessler_Vosegaard_2016}. (a) Picture of the modified probe showing the piezo motor and SC kit mounted. (b) Illustration of the main components of the SC NMR kit. The goniometer is from ref.~\citealp{Vosegaard_Langer_Daugaard_Hald_Bildse_Jakobsen_1996}. Adapted  from \cite{Vinding_Kessler_Vosegaard_2016} with permission.}
	\label{fig:sc_kit}
\end{figure*}

\begin{figure}[pos=t]
	\centering
		\includegraphics[width=\linewidth]{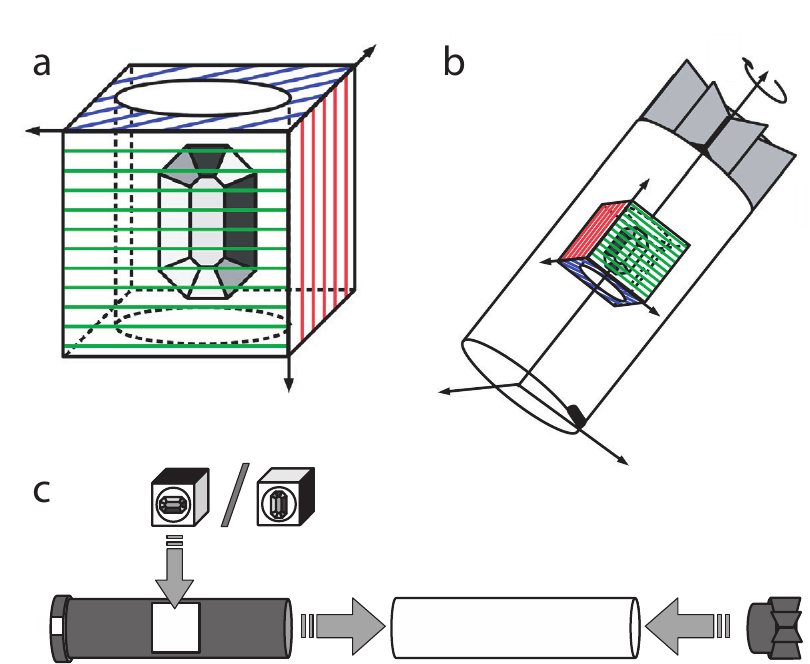}
	\caption{Sketches of the setup used for rotor-synchronised MAS SC NMR experiments \cite{Avadhut_2017}. (a) Hollow cube for mounting a single crystal, (b) cube and crystal as mounted in a MAS rotor, (c) polyoxymethylene rotor insert to accommodate the cube that fits into a 4 mm MAS rotor. Adapted  from \cite{Avadhut_2017} with permission.}
	\label{fig:mas_sc}
\end{figure}

While the requirement of dedicated equipment for SC NMR is a limitation for the use of this method, since only few laboratories have such equipment, some attempts have been made to modify existing NMR probes to allow for SC NMR measurements. Such modifications become increasingly relevant, as the RF side of NMR probes are more and more challenging to build as the RF frequencies increase. Hence, it is very convenient to use the RF side of an existing probe and introduce modifications only to the mechanical design. Preferably, such modifications should be reversible and simple to make, such that a probe with other functionalities can be used for SC NMR studies now and then.

Kov{\'a}cs and Rohonczy developed a SC NMR kit to modify a 7 mm MAS probe for SC NMR capabilities \cite{Kovacs_Rohonczy_2006} as shown in Fig.~\ref{fig:mas_probe}. The main idea was to change the rotation axis from the magic angle to a perpendicular orientation and to invoke the rotation by a cord connected to a ratchet wheel providing a precise stepwise rotation. The ratchet wheel was produced at a low cost by plexi glass, and the crystal was glued directly onto the wheel. The perpendicular orientation of the rotation axis was ensured by using a laser and mirror system, and similarly the stepwise rotation was calibrated using this system.

We recently used a similar idea but instead of using a MAS probe, we modified a static Bruker probe for SC NMR capabilities \cite{Vinding_Kessler_Vosegaard_2016}. The basic idea of this design, shown in Fig.~\ref{fig:sc_kit}, was to add a scaffold consisting of three brass rod extenders and a plate referred to as the probe interface in Fig.~\ref{fig:sc_kit}b. The plate is placed above the RF coil and serves as a platform onto which the goniometer may be attached. In order to remove the SC NMR kit, three screws attaching the three brass rods to the probe should be unscrewed, then the entire goniometer may be dismantled. In contrast to conventional SC NMR equipment, where the stepwise rotation is controlled from the bottom of the probe by a rod or string, this probe design used a small piezo-electric motor. A piezo-electric motor does not rely on magnetic parts and may therefore be mounted directly in the probe avoiding any gearing and installation of drive rod through the probe \cite{Vinding_Kessler_Vosegaard_2016}.

Avadhut et al.~\cite{Avadhut_2017} pursued a different approach to performing SC NMR studies by designing a crystal mounting device that fits into a 4 mm MAS rotor as illustrated in Fig.~\ref{fig:mas_sc}. Instead of accomplishing a stepwise rotation of the crystal and record NMR spectra for individual crystal orientations, they recorded carefully synchronized MAS spectra with different orthogonal mountings of the single crystal in the rotor. Clearly the filling factor is not good in this design, but due to the sample spinning lines are narrower than in conventional SC NMR, since MAS averages susceptibility broadening and dipolar interactions. An additional advantage of the approach is that no modifications are required to the MAS probe, it just requires that the MAS detector sends a trigger signal to the acquisition computer to allow rotor-synchronised initialisation of each experiment. It may be noted that this design could very nicely be combined with an inductively coupled microcoil \cite{Sakellariou_Goff_Jacquinot_2007} for sensitive detection of tiny crystals.

%%%%%%%%%%%%%%%%%%%%%%%%%%%%%%%%%%%%%%%%%
% ANALYSIS OF SC NMR DATA
%%%%%%%%%%%%%%%%%%%%%%%%%%%%%%%%%%%%%%%%%
\section{Analysis of Single-Crystal NMR Data}\label{sec:analysis}

The analysis of SC NMR is simple compared to the analysis of powdered samples, as no powder averaging is needed. Obviously, the description of the effect of the stepwise rotation of the crystal needs be considered through appropriate coordinate transformations as worked out below. Nonetheless, as the conventional SC NMR experiment involves only static NMR spectra, we may write up expressions for the resonance frequency depending on the crystal orientation, and in general we need not do any time propagation of the density operator.

\subsection{Theory}

In the high-field approximation, the resonance frequency for an NMR interaction is given by

\begin{equation}
  \nu_{\lambda,m} = \expval{m}{\H_\lambda}{m} - \expval{m-1}{\H_\lambda}{m-1}
\end{equation}

\noindent for observation of the $m \leftrightarrow m-1$ transition. $\lambda$ is the nuclear spin interaction, e.g. the quadrupole coupling, chemical shift, and hetero- or homonuclear dipole-dipole coupling interactions. In the secular approximation, a first-order interaction is given by

\begin{equation}
  \H_\lambda^{(1)} = (A_{00}^\lambda)^L \hat{T}_{00}^\lambda + (A_{20}^\lambda)^L \hat{T}_{20}^\lambda \label{eq:h1}
\end{equation}

\noindent with $(A_{j0}^\lambda)^L$ representing the spatial part of the $j$'th rank tensor expressed in the laboratory frame ($L$) and $\hat{T}_{j0}$ representing the spin part of the Hamiltonian. The second-order quadrupolar interaction is often relevant to consider. Its terms arise from the product of the second-rank tensor elements to yield

\begin{eqnarray}
  \H_Q^{(2)} &=& \frac{1}{2\nu_0} \left\{ 
   (A_{2-2}^Q)^L (A_{22}^Q)^L [2I(I+1) - 2\I_z^2 - 1] \I_z \right. \nonumber \\
   && +\left. (A_{2-1}^Q)^L (A_{21}^Q)^L [4I(I+1) - 8\I_z^2 - 1] \I_z \right\} \label{eq:hq2}
\end{eqnarray}

\begin{table*}[width=2\linewidth,cols=5,pos=h]
\caption{Spatial ($(A_{jm}^\lambda)^P$) and spin ($\hat{T}_{jm}^\lambda$) irreducible spherical tensor elements of the different nuclear spin interactions, $\lambda$: CS: chemical shift, J: J coupling (homo- or heteronuclear), IS: dipolar coupling (homo- or heteronuclear), and Q: quadrupolar coupling.}
\label{table:interactions}
\begin{tabular*}{\tblwidth}{@{} CCCCC@{} }
\toprule
Interaction, $\lambda$ & CS & J & IS & Q \\
\midrule
$A_{00}^\lambda$ & $\delta_\mathrm{iso}$ & $J_\mathrm{iso}$ & 0 & 0 \\
$(A_{20}^\lambda)^P$ & $\sqrt{\frac{3}{2}}\delta_\mathrm{aniso}$ & $\sqrt{\frac{3}{2}}J_\mathrm{aniso}$ & $\sqrt{\frac{3}{2}}b_{IS}$ & $\sqrt{\frac{3}{2}}C_Q$ \\
$(A_{2\pm 2}^\lambda)^P$ & $-\frac{1}{2}\delta_\mathrm{aniso} \eta_{CS}$ & $-\frac{1}{2}J_\mathrm{aniso} \eta_J$ & 0 & $-\frac{1}{2}C_Q \eta_Q$\\
\midrule
$\hat{T}_{00}^\lambda$ & $\nu_0 \I_z$ & $\hat{\boldsymbol{I}} \cdot \hat{\boldsymbol{S}}$ & 0 & 0 \\
$\hat{T}_{20}^\lambda$ & $\sqrt{\frac{2}{3}}\nu_0 \I_z$ & $\frac{1}{\sqrt{6}} (3\I_z \hat{S}_z- \hat{\boldsymbol{I}} \cdot \hat{\boldsymbol{S}})$ & $\frac{1}{\sqrt{6}} (3\I_z \hat{S}_z- \hat{\boldsymbol{I}} \cdot \hat{\boldsymbol{S}})$ & $\frac{1}{\sqrt{6}} (3\I_z^2 - I(I+1))$ \\
\bottomrule
\end{tabular*}
\end{table*}

\noindent The laboratory-frame representation, $(A_{jm}^\lambda)^L$, is obtained through a series of coordinate transformations starting in the principal axis frame ($P$) through the crystal holder frame ($C$) to the laboratory frame ($L$). The $(A_{jm}^\lambda)^P$ elements in the principal axis frame and the $\hat{T}_{jm}^\lambda$ elements are given in Table \ref{table:interactions}. 

The coordinate transformations are typically carried out using three Euler angles, which for a transformation from frame $A$ to frame $B$ first perform a rotation around the $z_A$ axis, then a rotation around the new $y$ axis that rotates $z_A$ into $z_B$, and finally a rotation around the $z_B$ axis. The cartesian rotation matrices the rotation around the $z$ axis is given by

\begin{equation}
  R_{ez}(\alpha) = \left(\begin{array}{ccc}
  \ca & \sa & 0 \\
  -\sa & \ca & 0 \\
  0 & 0 & 1
  \end{array}\right)
  \label{eq:rez}
\end{equation}

\noindent where $c_a = \cos \alpha$ and $s_\alpha = \sin \alpha$. The rotation around the $y$ axis is given by

\begin{equation}
  R_{ey}(\beta) = \left(\begin{array}{ccc}
  \cb & 0 & -\sb \\
  0 & 1 & 0 \\
  \sb & 0 & \cb 
  \end{array}\right)
  \label{eq:rey}
\end{equation}

\noindent The total rotation matrix for the three consecutive rotations transforming from frame $A$ to $B$ is given as

\begin{eqnarray}
  R_{AB}(\alpha, \beta, \gamma) = R_{ez}(\gamma) R_{ey}(\beta) R_{ez}(\alpha) \hspace{0.3\linewidth} \nonumber \\
  =
  \left(\begin{array}{ccc}
  \ca \cb\cg - \sa\sg & \sa\cb\cg + \ca\sg & -\sb\cg \\
  -\ca\cb\sg - \sa\cg & -\sa\cb\sg + \ca\cg & \sb\sg \\
  \ca\sb & \sa\sb & \cb
  \end{array}\right)
  \label{eq:rab}
\end{eqnarray}

\begin{figure}[pos=t]
	\centering
		\includegraphics[width=0.7\linewidth]{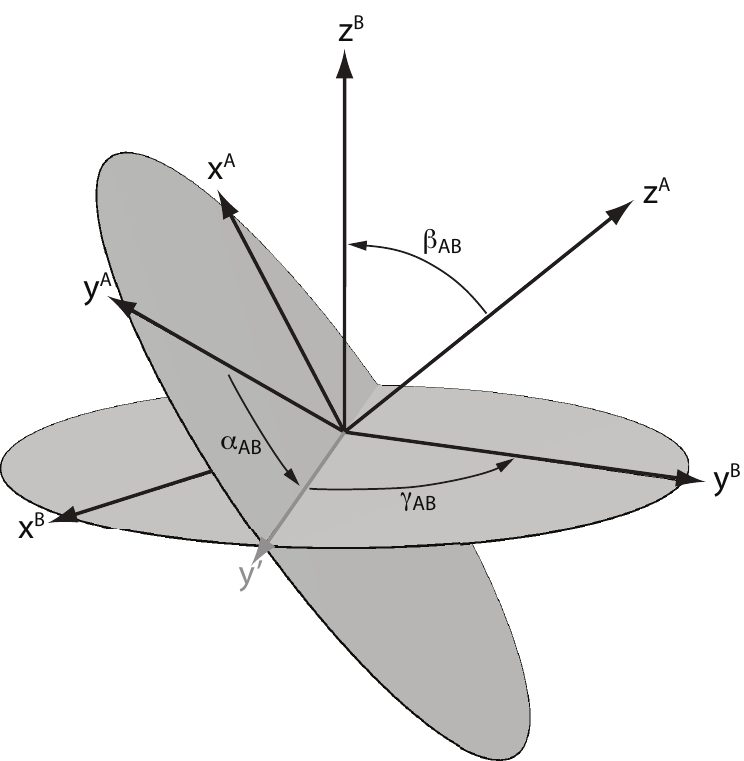}
	\caption{Illustration of the rotations through three Euler angles ($\alpha_\mathrm{AB}, \beta_\mathrm{AB}, \gamma_\mathrm{AB}$) that allow one to relate the two frames marked A and B.}
	\label{fig:euler}
\end{figure}

\noindent $\alpha$ performs a rotation around the $z$ axis of frame $A$, $\beta$ performs a rotation around the new $y$ axis, and $\gamma$ represents a rotation around the $z$ axis of frame $B$. Figure \ref{fig:euler} displays these three steps and the resulting orientations of the coordinate systems.

While it is in general most convenient to consider the Hamiltonians as irreducible spherical tensors, most single crystal studies have described the Hamiltonian in Cartesian space, since this relates directly to the rotations. For the rotation around the $x$, $y$, and $z$ axes of the crystal holder system, the rotation matrices describing the $C\rightarrow L$ transformations become

\begin{equation}
R_x(\theta) = 
 \left(\begin{array}{ccc}
  1 & 0 & 0 \\
  0 & \cos\theta & \sin\theta \\
  0 & -\sin\theta & \cos\theta
  \end{array}\right)
  \label{eq:rx}
\end{equation}
\begin{equation}
R_y(\theta) = 
 \left(\begin{array}{ccc}
  0 & -1 & 0 \\
  \cos\theta & 0 & \sin\theta \\
  -\sin\theta & 0 & \cos\theta
  \end{array}\right)
  \label{eq:ry}
\end{equation}
\begin{equation}
R_z(\theta) = 
 \left(\begin{array}{ccc}
  0 & 0 & 1 \\
  \sin\theta & -\cos\theta & 0 \\
  \cos\theta & \sin\theta & 0
  \end{array}\right)
  \label{eq:rz}
\end{equation}

\noindent Note that these matrices apply to the crystal tenon system shown in Fig.~\ref{fig:vosegaard3} and are the result of nesting of several Euler matrices, so they cannot easily be described in the form of Eq.~\ref{eq:rab}. For other SC probe setups the actual forms of Eqs.~\ref{eq:rx}-\ref{eq:rz} may need be revised to reflect the geometry of the particular probe/goniometer.

The actual rotation of the Hamiltonian through the $P \rightarrow C \rightarrow L$ cascade of transformations is given by

\begin{equation}
  (\H_\lambda)^{q,L} = R_q R_{PC} (\H_\lambda)^{q,P}R_{PC}^T R_q^T, \label{eq:htrans}
\end{equation}

\noindent where the index $q$($=x$, $y$,or $z$) refers to the mounting of the crystal, the index $P/L$ refers to the frame of reference, and the superscript $T$ indicates the transposed matrix. Combining the expression in Eq.~\ref{eq:htrans} with the frequency expressions in Eqs.~\ref{eq:h1} and \ref{eq:hq2}, we obtain the generalised expressions for the resonance frequency

\begin{eqnarray}
  \nu_{\lambda,q}^{(1)} & = & A_{\lambda,q}^{(1)} + B_{\lambda,q}^{(1)} \cos2\theta + C_{\lambda,q}^{(1)} \sin2\theta \label{eq:nu1} \\
  \nu_{\lambda,q}^{(2)} & = & A_{\lambda,q}^{(2)} + B_{\lambda,q}^{(2)} \cos\theta + C_{\lambda,q}^{(2)} \sin\theta \nonumber \\
  & & \hspace{1cm} + D_{\lambda,q}^{(2)} \cos2\theta + E_{\lambda,q}^{(2)} \sin2\theta \label{eq:nu2}
\end{eqnarray}

\noindent We note the general dependency on $2\theta$ (and $4\theta$ for second-order interactions) which is characteristic for SC NMR setups where the rotation axis is perpendicular to the magnetic field. If the rotation angle is different from $90^\circ$, terms involving $\theta$ (and $3\theta$) also appear \cite{Vosegaard_Hald_Daugaard_Jakobsen_1999}. The expressions for the coefficients $M_{\lambda,q}$ ($M=A-E$) are given elsewhere \cite{Vosegaard_Langer_Daugaard_Hald_Bildse_Jakobsen_1996, Blinc_1991, Vosegaard_Skibsted_Bildse_Jakobsen_1996a, Volkoff_1953}.

\subsection{Crystal Symmetry}\label{sec:symmetry}

It is always relevant to know the orientation of the nuclear spin interaction tensors in the molecular frame of reference. In solid state NMR, this is possible through experiments determining the relative orientation of e.g. a chemical shift tensor and one or more dipolar tensors, as the latter interactions may directly be related to the molecular structure. An interesting feature of SC NMR is the ability to directly determine the orientation of the nuclear spin interaction tensors in the molecular frame through determination the orientation of the crystallographic axes and the nuclear spin interaction tensors in the crystal holder frame. Kennedy and Ellis investigated the implications of different crystal symmetries and how appropriate transformations and possible orthogonalization could be done \cite{Kennedy_Ellis_1989a}. The orientation of the crystallographic axes are typically determined either by accompanying X-ray diffraction studies of the same crystal or by indexing the crystal faces by optical microscopy.

Kennedy and Ellis \cite{Kennedy_Ellis_1989a} also investigated the role of the local symmetry at the position of the atom studied by NMR, which imposes constraints on the nuclear spin interactions for this nucleus. For example, if a nucleus is located on a two-fold rotation axis, this would imply that its nuclear spin interaction tensors should be identical when applying this symmetry element. Let us consider the (symmetric) chemical shift tensor

\begin{equation}
   \s = \left(\begin{array}{ccc}
   \s_{xx} & \s_{xy} & \s_{xz} \\
   \s_{xy} & \s_{yy} & \s_{yz} \\
   \s_{xz} & \s_{yz} & \s_{zz}
  \end{array}\right)
  \label{eq:sigma}
\end{equation}

\noindent that experiences a two-fold rotation around the $z$ axis corresponding to the following rotation matrix

\begin{equation}
   R_2 = \left(\begin{array}{ccc}
  -1 & 0 & 0 \\
  0 & -1 & 0 \\
  0 & 0 & 1
  \end{array}\right)
  \label{eq:rtwo}
\end{equation}

\noindent The transformed tensor is given by

\begin{equation}
  \s' = R \s R^T = \left(\begin{array}{ccc}
   \s_{xx} & \s_{xy} & -\s_{xz} \\
   \s_{xy} & \s_{yy} & -\s_{yz} \\
   -\s_{xz} & -\s_{yz} & \s_{zz}
  \end{array}\right),
\end{equation}

\noindent and since $\s = \s'$ this implies that $\s_{xz} = -\s_{xz} = 0$ and $\s_{yz} = -\s_{yz} = 0$
leading to the shift tensor

\begin{equation}
   \s = \left(\begin{array}{ccc}
   \s_{xx} & \s_{xy} & 0 \\
   \s_{xy} & \s_{yy} & 0 \\
   0 & 0 & \s_{zz}
  \end{array}\right).
  \label{eq:sigma2}
\end{equation}

\noindent This shows us that a nuclear spin interaction tensor associated with a nucleus located on a two-fold axis must have one of its principal elements aligned along the two-fold axis, but without any constraints on the other elements. If the symmetry element was a mirror plane, one of the principal elements should be aligned with the mirror plane normal.

Higher symmetry axes, e.g. three- or four-fold axes lead to the further constraint that the nuclear spin interaction tensor must be axially symmetric with the unique principal element aligned along the symmetry axis.

Since the NMR experiment is not affected by translation, screw axes have the same affect as the corresponding rotational symmetry element and glide planes have the same effect as mirror planes.

Given the above consequences of the local symmetry at the position of the observed nucleus as well as the overall crystal symmetry on the nuclear spin interaction tensors, it was rather surprising to read the paper by Kim and co-workers \cite{Kim_Lee_Kim_2016} reporting on a symmetry-breaking quadrupole coupling tensor for $^{133}$Cs in Cs$_2$CrO$_4$. The author's obscure assertion was that the Cs nuclei are floating in space and need not respect the crystal symmetry. Considering that the forces placing the atoms and thereby also the nuclei in a crystal arise from covalent bonds and electrostatic interactions between atoms, which are arranged following the crystal symmetry, this argument does not seem to fully justify the observed deviation from the crystal symmetry, determined by XRD \cite{Miller_1938, Aleksovska_Petruevski_Pejov_1997} and later backed up by neutron diffraction \cite{Morris_Kennard_Moore_Smith_Montgomery_1981}. A previous SC NMR study of Wasylishen and co-workers \cite{Power_Mooibroek_Wasylishen_Cameron_1994} determined the orientation of both the chemical shift and quadrupole coupling tensors to be consistent with the crystal structure, and our MAS study \cite{Skibsted_Vosegaard_Bildsoe_Jakobsen_1996} also confirmed this consistency. Finally, a recent theoretical study confirmed the observation of tensor orientations respecting the local crystal symmetry \cite{czernek_describing_2017}.

\subsection{Software for Analysis of Single-Crystal NMR Data}

The process of analysing the spectra from a conventional three-axis (or two-axis) SC NMR probe involves four steps
\begin{enumerate}
\item Peak picking to generate lists of peaks in the individual spectra.
\item Assignment of the peaks for different rotation angles for the same crystal mounting in rotation plots.
\item Combining rotation plots for different crystal mountings to establish the correlation between these.
\item Fitting the resonance positions for all crystal mountings and -orientations to provide a final set of nuclear spin interaction parameters.
\end{enumerate}

\begin{figure}[pos=t]
	\centering
		\includegraphics[width=\linewidth]{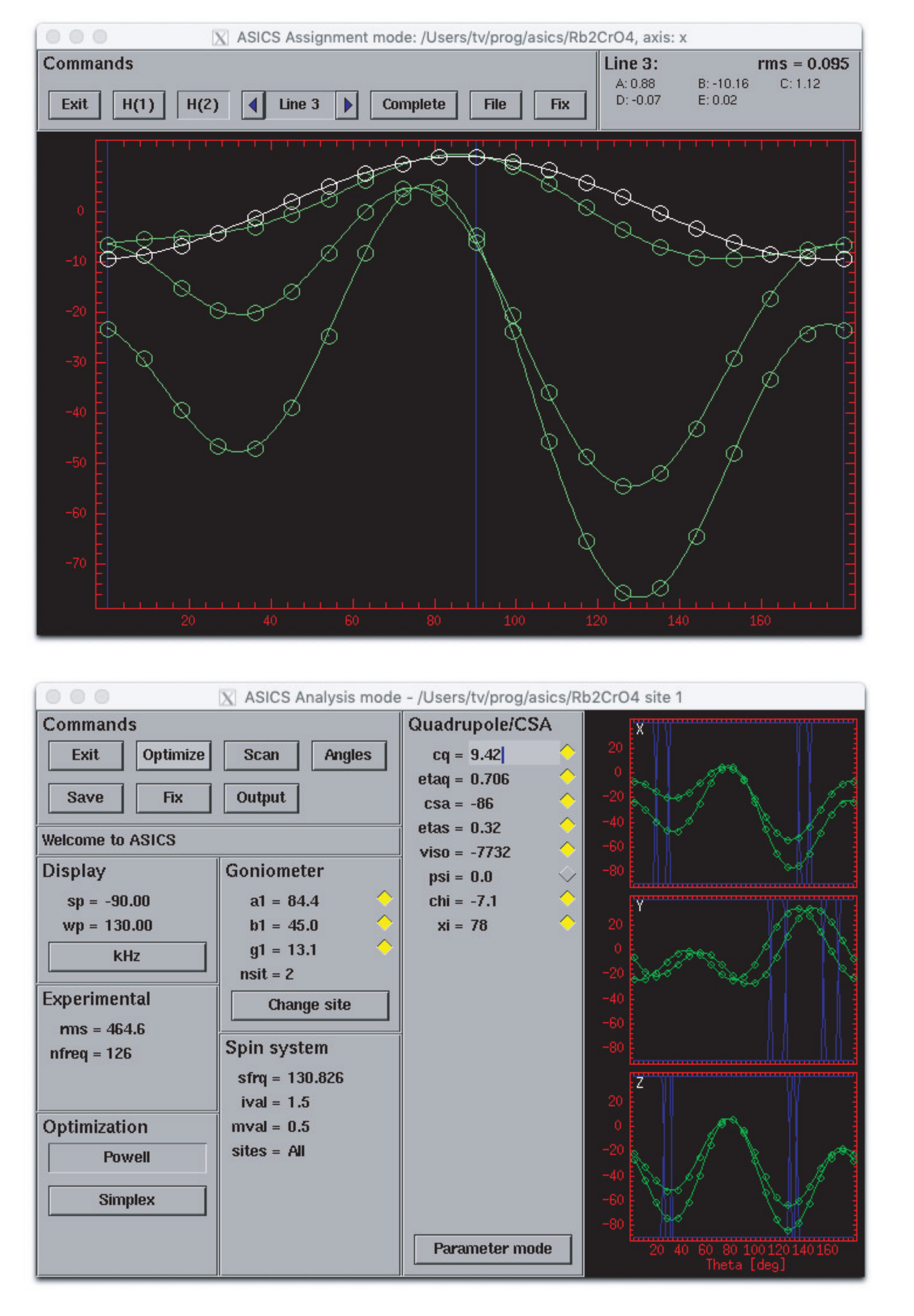}
	\caption{Screenshots of displays generated by the ASICS software package for analysis of SC NMR data. The top figure shows the window for assigning resonances in the individual rotation plots, and the bottom figure shows the window for the fitting the nuclear spin interaction parameters to the full sets of rotation plots.}
	\label{fig:asics}
\end{figure}

\noindent Since these steps may be automated to a certain degree, software tools have been developed in different laboratories to deal with the various processes, but very little of this work has been published. The software tools Analysis of Single-Crystal Spectra (ASICS) \cite{Vosegaard_Hald_Langer_Skov_Daugaard_Bildse_Jakobsen_1998} was created to help performing these four points. The peak picking (step 1) may be done by deconvoluting through manual peak-picking of lines for the individual spectra, which generates unassigned lists of resonances for each rotation angle. These lists (for each rotation axis) are assigned to individual curves (step 2) described by the appropriate functions for the angular dependence of the resonance frequency (e.g. Eqs.~\ref{eq:nu1}-\ref{eq:nu2}) as illustrated in Fig.~\ref{fig:asics} (top panel). Following this, matching curves for the different rotation axes are combined (step 3) to obtain a first estimate of the nuclear spin interaction parameters as described in ref.~\citealp{Vosegaard_Hald_Langer_Skov_Daugaard_Bildse_Jakobsen_1998}. Finally, the nuclear spin interaction parameters are fitted to the experimentally determined resonance positions through least-squares optimization (step 4), in ASICS using the graphical user interface shown in Fig.~\ref{fig:asics} (bottom panel).

While the ASICS package was developed in the late 90'es and thus is more than 20 years old, it is still in operation on modern computers and available for download (https://nmr.au.dk), but it is not likely to be further developed as it relies on outdated graphics libraries. 

\begin{figure}[pos=t]
	\centering
		\includegraphics[width=0.8\linewidth]{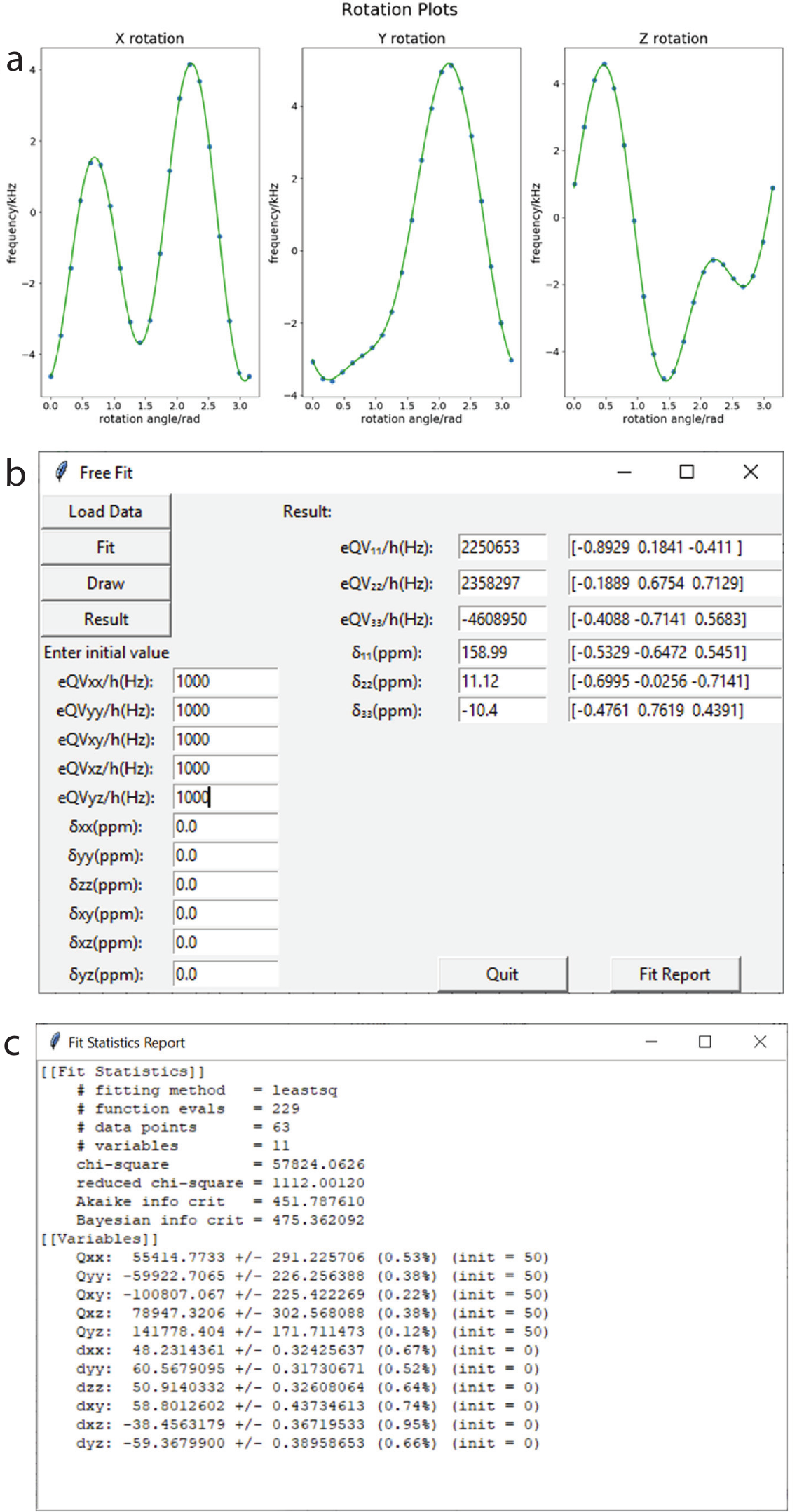}
	\caption{Screenshots of displays generated by the SCFit program showing (a) rotation plots with experimental frequencies as black dots and theoretical curves as green lines, (b) the parameter window showing all parameters for the model, and (c) the fit status window reporting various information about the fit of the rotation plots. Adapted  from  \cite{Xu_Bryce_2019} with permission.}
	\label{fig:scfit}
\end{figure}

Xu and Bryce recently presented the Python-based \\ (https://www.python.org) SCFit software package \cite{Xu_Bryce_2019} that was developed to allow interpretation of all nuclear spin interactions including dipolar and spin-spin couplings in addition to the chemical shift and quadrupolar coupling. This program is more versatile than ASICS and allows fitting tensors and their orientation directly. Screenshots of the fitting window and rotation plots are shown in Fig.~\ref{fig:scfit}, and the program is available from the author's web site \\ (https://mysite.science.uottawa.ca/dbryce/software.html).

\begin{figure}[pos=t]
	\centering
		\includegraphics[width=\linewidth]{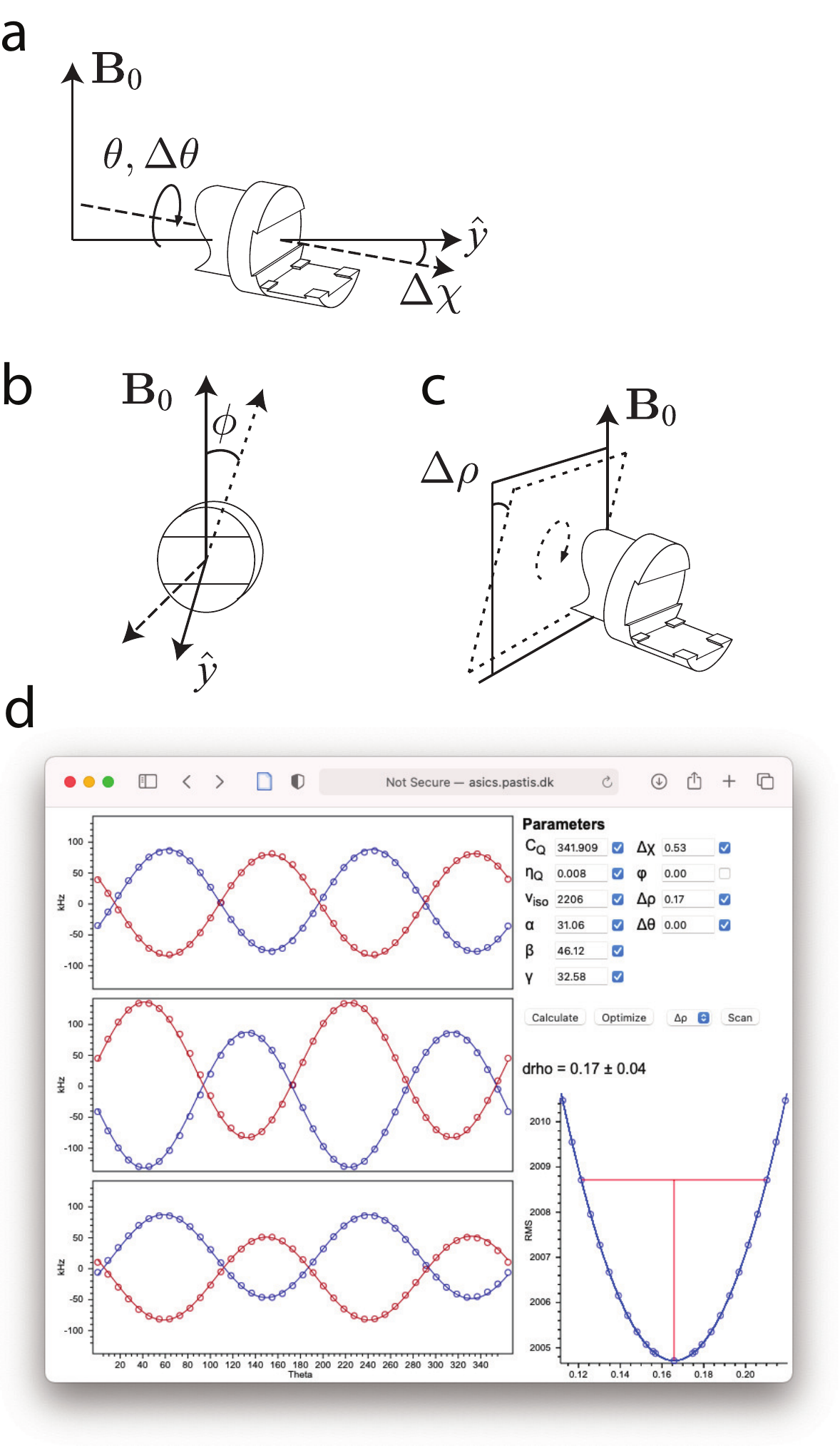}
	\caption{(a-c) Sketches showing possible deviations from an ideal geometry of a SC NMR goniometer setup. The angles $\Delta \chi$ (a) and $\phi$ (b) describe a misalignment of the goniometer with respect to the axis of rotation, and the angle $\Delta\rho$ (c) describes the deviation of the rotation axis from the desired orientation perpendicular to the magnetic field direction. (d) Screenshot of a display generated by the webASICS application showing rotation plots with corresponding fits of a quadrupolar tensor. The lower right corner shows a 95\% confidence-interval calculation in the graph displaying the $\chi^2$ value as a function of the angle $\Delta \rho$.}
	\label{fig:webasics}
\end{figure}

A few years ago, we developed a SC NMR goniometer device to mount on a conventional static probe (see Fig.~\ref{fig:sc_kit}). One of the trade-offs of this probe is that the normally precise geometry of SC NMR probes (e.g. the crystal rotation axis is precisely perpendicular to the magnetic field) may be compromised. Hence, we needed to revise the analysis of the SC NMR data obtained with this goniometer design to include these deviations in the analysis. Based on our recent experiences with development of web-based NMR software \cite{Vosegaard_2015,Vosegaard_2018a,JUHL20201}, we decided to implement this type of analyses in the webASICS \cite{Vinding_Kessler_Vosegaard_2016} software. Figure \ref{fig:webasics}a-c shows the angular deviations possibly encountered with the SC goniometer kit. It turns out that the SC NMR measurements, relying on frequencies rather than the intensity distribution observed for powder samples, provides sufficient data quality to independently determine these parameters as illustrated by the $\chi^2$ fit shown in the lower right corner of the webASICS screen shot in Fig.~\ref{fig:webasics}, where the value for the angular deviation $\Delta\rho$ has been determined to $0.17^\circ \pm 0.04^\circ$, with the error bar representing the 95\% confidence interval. 

We found it very encouraging that it is possible to determine the angular deviations from the ideal geometry with such high precision. In the future, we recommend to use this model to fit experimental SC NMR data, as minor imperfections in SC NMR data have often been observed in the literature, which may be attributed to small deviations and uncertainties in the crystal mounting/SC NMR probe design.

A number of studies have followed a different approach to the analysis of SC NMR data. Instead of using dedicated SC NMR analysis software, the versatile simulation program for solid-state NMR (SIMPSON) \cite{simpson_2000, simpson_2014} has been used to analyse such data \cite{Hansen_Vosegaard_Jakobsen_Skibsted_2004, Nieuwendaal_Mattler_Bertmer_Hayes_2011, Avadhut_2017}. Some of these will be addressed below.

\subsection{Assignment Challenges}

It may be challenging to correlate the rotation plots for two or three different rotation axes. With the conventional three-axis goniometer setup this assignment will be assisted by the fact that some of the different mountings of the crystal will lead to the same orientation of the crystal relative to the magnetic field

\begin{eqnarray}
\mathrm{Mounting}\ 1\ (\theta = 0^\circ) & \equiv & \mathrm{Mounting}\ 2\ (\theta = 0^\circ) \label{m:xy} \\
\mathrm{Mounting}\ 1\ (\theta = 90^\circ) & \equiv & \mathrm{Mounting}\ 3\ (\theta = 90^\circ) \phantom{xxxx} \\
\mathrm{Mounting}\ 2\ (\theta = 90^\circ) & \equiv & \mathrm{Mounting}\ 3\ (\theta = 0^\circ) \label{m:yz}
\end{eqnarray}

\noindent where the different mountings are shown in Fig.~\ref{fig:vosegaard3}c-\ref{fig:vosegaard3}d. The first case (Eq.~\ref{m:xy}) is easily understood, since these two mountings only differ by a $90^\circ$ rotation around the magnetic field direction. Likewise, the other cases may also be realised by following the definition of the rotations and mountings in Fig.~\ref{fig:vosegaard3}. Two-axis goniometers or other goniometer designs may have similar mountings that lead to identical spectra. For example, our two-axis goniometer leads to identical spectra if the crystal is mounted in the (a) and (b) mountings (cf.~Fig.~\ref{fig:vosegaard2}d) at rotation angles of $\theta_{(a)} = \theta_\mathrm{eq}$ and $\theta_{(b)} = 360^\circ -\theta_\mathrm{eq}$, with $\theta_\mathrm{eq} = \arccos(\tan \phi)$ and $\phi$ defined in Fig.~\ref{fig:vosegaard2}c \cite{Vosegaard_Hald_Daugaard_Jakobsen_1999}.

It is our experience that mounting the crystal in a random orientation makes the assignment easier than if the crystal is mounted with one of its crystallographic axes along one of the goniometer axes, especially for crystals with high symmetry. Indeed, magnetically inequivalent but crystallographically equivalent sites often result in the same resonance frequencies when oriented with one of the symmetry axes aligned along the magnetic field direction, leading to overlap of their resonances in the NMR spectrum. In the case that the crystal is mounted with its crystallographic axes along the tenon system (Fig.~\ref{fig:vosegaard3}d), these orientations may coincide with the six mountings listed in Eqs.~\ref{m:xy}-\ref{m:yz}, and hence we lose the ability to use these constraints for the assignment.

There are other parts of the assignment that may be challenging. There are many cases where SC NMR has been used to determine both the quadrupole coupling and chemical shift tensors for quadrupolar nuclei with half-integer spin. In cases of relatively weak quadrupole couplings, where it is possible to observe the satellite transitions, the quaderupole coupling tensor is easily determined with high precision from the satellite transitions, while the chemical shift tensor is conveniently determined from the central transition, where the often dominant $\H_Q^{(1)}$ term is not present. In such cases, if several different sites or magnetically inequivalent nuclei are present, it may be difficult to correlate the satellite transitions with the correct central transition for each site. 

We have pursued the problem of correlating the resonances for the satellite transitions to the correct central transition by two different approaches:

\begin{figure}[pos=t]
	\centering
		\includegraphics[width=\linewidth]{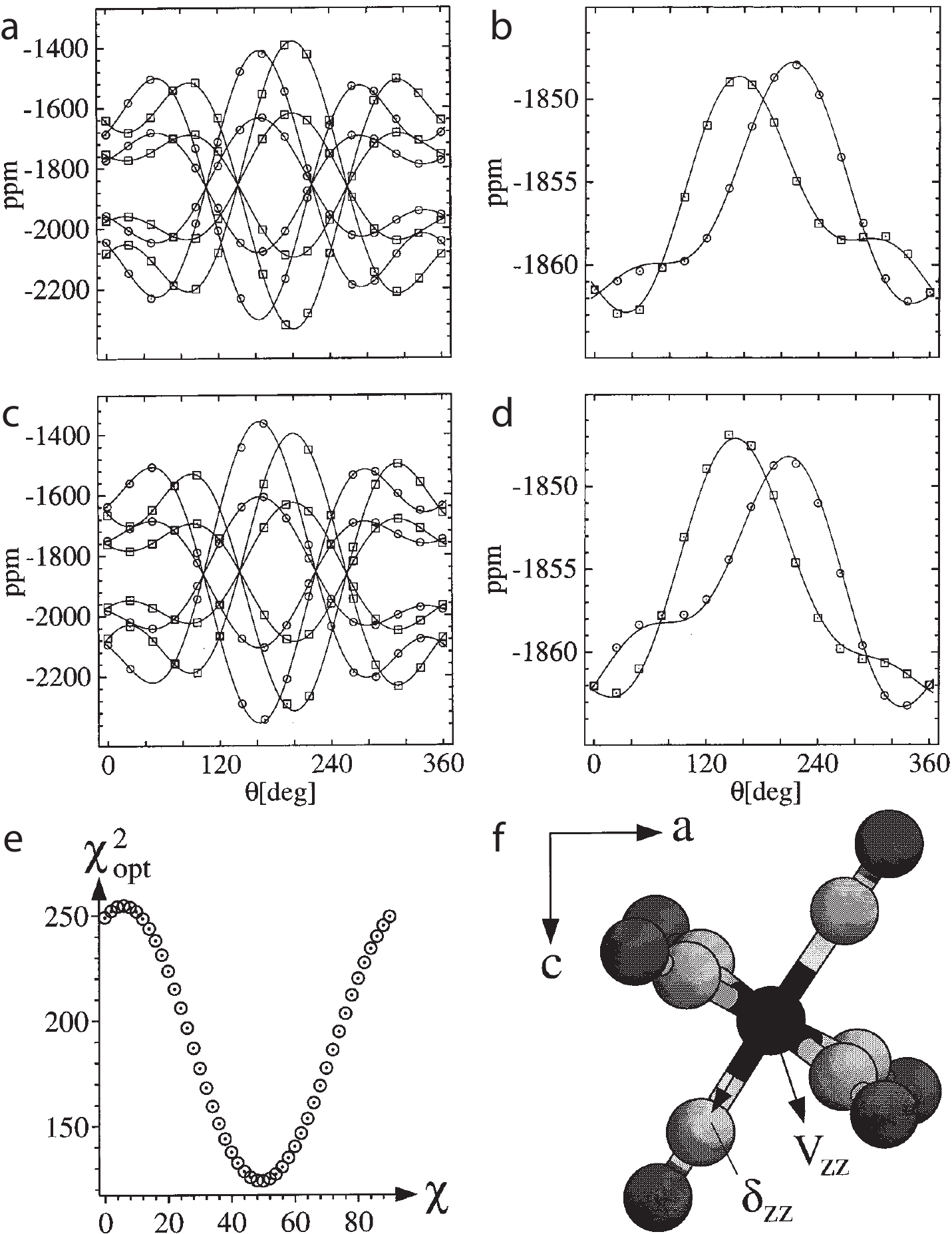}
	\caption{SC NMR rotation plots of the resonance frequencies of $^{95}$Mo in Mo(CO)$_6$  for the satellite (a,c) and central (b,d) transitions for the a (a,b) and b (c,d) mounting of our two-axis goniometer (cf.~Fig.~\ref{fig:vosegaard2}). (e) $\chi^2$ deviation between experimental and simulated resonance frequencies for the satellite transitions for different values of the angle $\chi$ describing the relative orientation of the quadrupole coupling and chemical shift tensors. (f) orientation of the two tensors in the crystal frame. Adapted from \cite{Vosegaard_Skibsted_Jakobsen_1999a} with permission.}
	\label{fig:95Mo}
\end{figure}

First, in a $^{95}$Mo study of Mo(CO)$_6$, the rotation plot of the satellite transitions (Figs.~\ref{fig:95Mo}a and \ref{fig:95Mo}c) revealed a relatively small quadrupole coupling of  $C_\mathrm{Q} = 89.3$ kHz. With this size of quadrupole coupling, it was clear that there would be no observable second-order quadrupole effect on the central transition (Figs.~\ref{fig:95Mo}b and \ref{fig:95Mo}d), and hence it would be difficult to correlate the satellite and central transitions based on the quadrupolar tensor \cite{Vosegaard_Skibsted_Jakobsen_1999a}. The central transition in this case was only sensitive to the chemical shift, but the satellite transitions, although dominated by the first-order quadrupolar interaction, would also be affected by the chemical shift interaction. The expected precision of the measurement of the chemical shift from the satellite transitions would be lower than from the central transition. With two crystallographically equivalent but magnetically inequivalent sites, there are two different possible assignments, and hence the two different possible assignments would lead to different relative orientations of the chemical shift and quadrupole coupling tensors. Since the Mo atom is located on a mirror plane, the relative orientations of the chemical shift and quadrupole coupling tensors may be defined by a single Euler angle ($\chi$, corresponding to the angle $\beta_\mathrm{AB}$ in Fig.~\ref{fig:euler}). In this study, we scanned the $\chi$ angle to search for the lowest $\chi^2$ deviation between the experimental resonances for the satellite transitions and the simulated frequencies and found a clear minimum around $\chi \approx 50^\circ$. The analysis of the central transition led to two minima for the two possible assignments. The incorrect assignment gave a value of $\chi \approx 14^\circ$, while what turned out to be the correct assignment gave a value of $\chi = 49.6^\circ \pm 0.8^\circ$ \cite{Vosegaard_Skibsted_Jakobsen_1999a}.

\begin{figure}[pos=t]
	\centering
		\includegraphics[width=\linewidth]{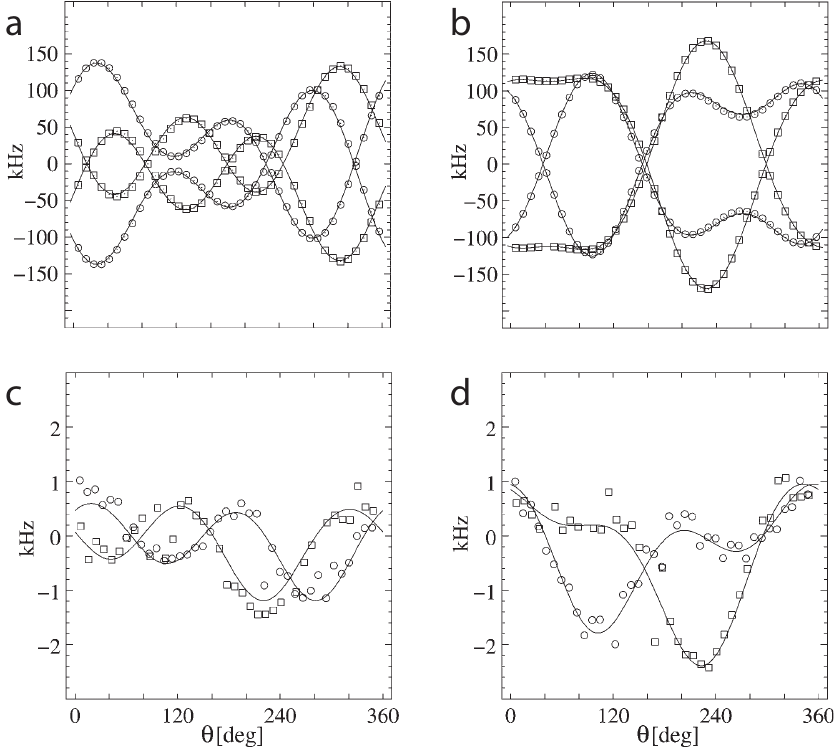}
	\caption{Rotation plots of the $^{11}$B SC NMR resonance frequencies from the satellite transitions for danburite (CaB$_2$Si$_2$O$_8$) for rotation about (a) the a axis and (b) the b axis of our two-axis SC probe \cite{Vosegaard_Hald_Daugaard_Jakobsen_1999}. (c,d) Rotation plots for the sum of the frequencies for the $m = \pm 3/2 \leftrightarrow \pm 1/2$ transitions for rotation about the a (c) and b (d) axes. Reproduced from \cite{Vosegaard_Hald_Daugaard_Jakobsen_1999} with permission.}
	\label{fig:11b}
\end{figure}

Second, in a $^{11}$B SC, MAS, and MQMAS NMR study of borates, we had recorded $^{11}$B SC NMR data for a number of borates with the aim of determining both their quadrupole-coupling and chemical-shift tensors. The quadrupole coupling tensor dominates the satellite transitions, which makes it difficult to obtain precise determination of the chemical shift from these transitions by direct fitting. To get the most precise values for the chemical shift from the satellite transitions, we chose to plot the sum of the frequencies for the resonances of the $m = \pm 3/2 \leftrightarrow \pm 1/2$ and  transitions. Since the sign of $\H_Q^{(1)}$ changes between these transitions, the effect of the first-order quadrupolar Hamiltonian is eliminated by this approach. Removal of this dominant interaction allowed for accurate determination of the chemical shift parameters. However, this kind of analysis was not available in any dedicated SC NMR software, hence we chose to implement the SC fit in SIMPSON \cite{Hansen_Vosegaard_Jakobsen_Skibsted_2004}, where the individual rotations were calculated as different crystallites in the SIMPSON interface. Figure \ref{fig:11b} shows examples of the rotation plots for the $^{11}$B satellite transitions and sum of the satellites for danburite (CaB$_2$Si$_2$O$_8$) for the two rotation axes of our two-axis SC NMR probe (cf.~Fig.~\ref{fig:vosegaard2}) \cite{Vosegaard_Hald_Daugaard_Jakobsen_1999}.

\subsection{Reducing the Number of Rotation Axes}\label{sec:one}

Acquisition of full rotation plots for three different mountings of the crystal  requires long experiment time. In many cases, data from three rotation axes provide redundant information for determining the magnitude and orientation of the nuclear spin interactions. Just considering that the chemical shift tensor in its symmetric form is characterized by six parameters (three principal elements and three Euler angles describing the tensor orientation), and the rotation plots from each rotation axis with the rotation axis perpendicular to the magnetic field is described by three parameters (cf.~Eq.~\ref{eq:nu1}) suggests that two axes of rotation should be enough.

Shekar et al.~\cite{Shekar_Ramamoorthy_Wittebort_2002} demonstrated that it is possible simplify the SC NMR measurements by only requiring a single rotation axis when aiming to determine a chemical shift tensor. This is possible if the principal elements of the shift tensor are determined by other approaches, e.g. MAS or static powder NMR. This work was based on similar early work by Weil \cite{Weil_1973} and more recent work by Haeberlen and co-workers \cite{Tesche_Zimmermann_Poupko_Haeberlen_1993} who studied SC samples with special mountings of the crystal. 

Br\"auniger and co-workers \cite{Zeman_Hoch_Hochleitner_Bruniger_2018} elaborated on this approach and took advantage of the presence of chemically equivalent but magnetically inequivalent \vv\ and $^{207}$Pb nuclei to provide sufficient information about the nuclear spin interaction tensors from a single rotation axis. Indeed, the success of such studies depends on the actual crystal symmetry and local symmetry of the observed nuclei, and in some cases additional information about the orientation of the nuclear spin interaction tensors obtained by density-functional theory calculations were required \cite{Zeman_Hoch_Hochleitner_Bruniger_2018}.

Harbison and co-workers \cite{Kye_Zhao_Harbison_2005} used the line crossings of the NMR resonances from different crystallographically equivalent but magnetically inequivalent sites to work out the orientation of the crystallographic axes relative to the SC NMR setup and to determine the ful tensor information from a single rotation plot.

\begin{figure}[pos=t]
	\centering
		\includegraphics[width=\linewidth]{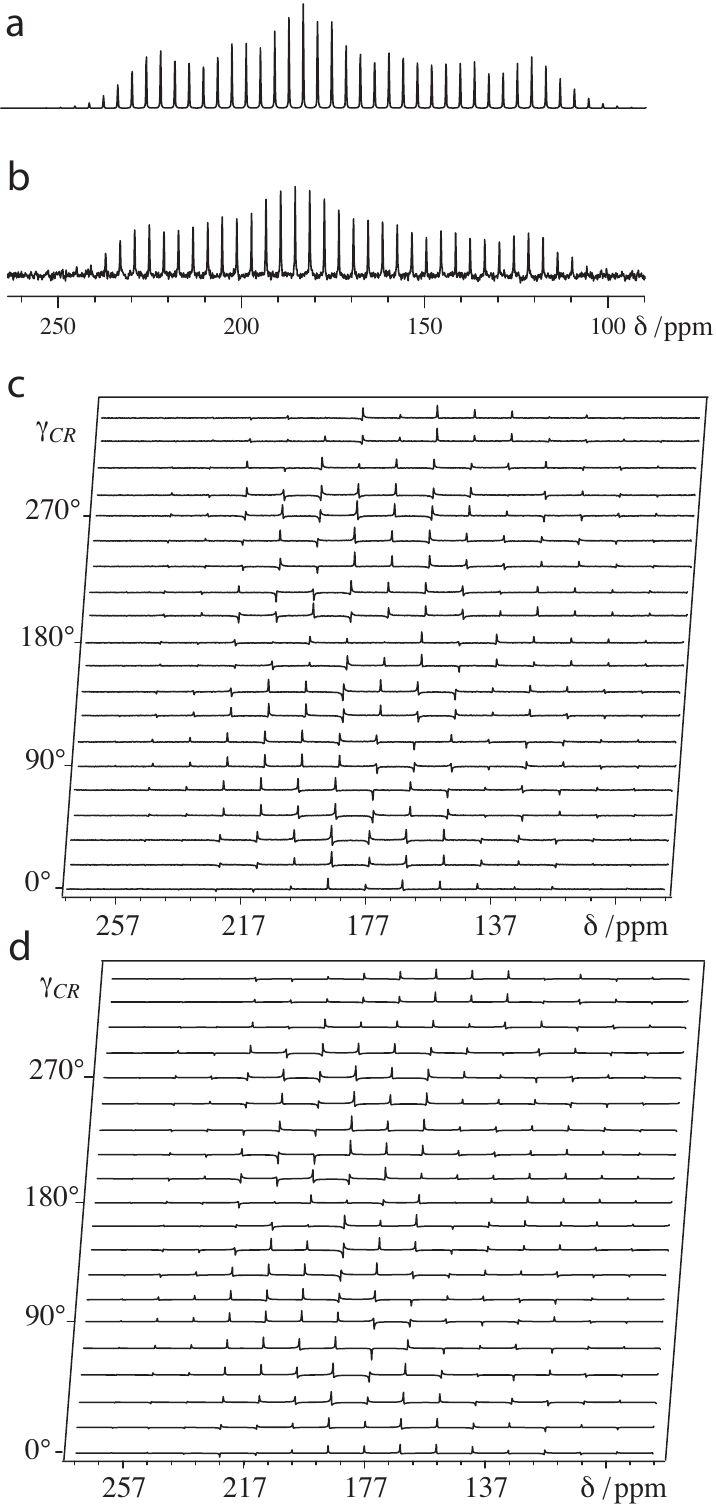}
	\caption{Expansions of the carbonyl region of $^{13}$C CPMAS NMR spectra of $L$-alanine. Experimental (b) and simulated (a) spectra of a powdered sample. Experimental (c) and simulated (d) spectra of a single crystal using a special SC MAS NMR setup. Adapted  from  \cite{Avadhut_2017} with permission.}
	\label{fig:mas_analysis}
\end{figure}

Avadhut et al.~\cite{Avadhut_2017} exploited the use of MAS for single crystals as already addressed in Fig.~\ref{fig:mas_sc}. The SC MAS NMR setup allowed two orthogonal mountings of the crystal (Fig.~\ref{fig:mas_sc}c). Figure \ref{fig:mas_analysis} shows the use of this approach to determine the $^{13}$C chemical shift tensors for the carbonyl in $L$-alanine. To determine the orientation of the chemical shift tensor, a powder spectrum was first recorded from which it is possible to determine the principal axis values (Figs.~\ref{fig:mas_analysis}a-\ref{fig:mas_analysis}b). MAS spectra of a single crystal were then recorded with different rotor pitch angles, represented by the angle $\gamma_\mathrm{CR}$ in Figs.~\ref{fig:mas_analysis}c-\ref{fig:mas_analysis}d. The different phases of the spinning sidebands for different rotor pitch angles reflect the sensitivity of the method towards the actual orientation of the $^{13}$C chemical shift tensor relative to the MAS rotor.

\begin{figure}[pos=t]
	\centering
		\includegraphics[width=\linewidth]{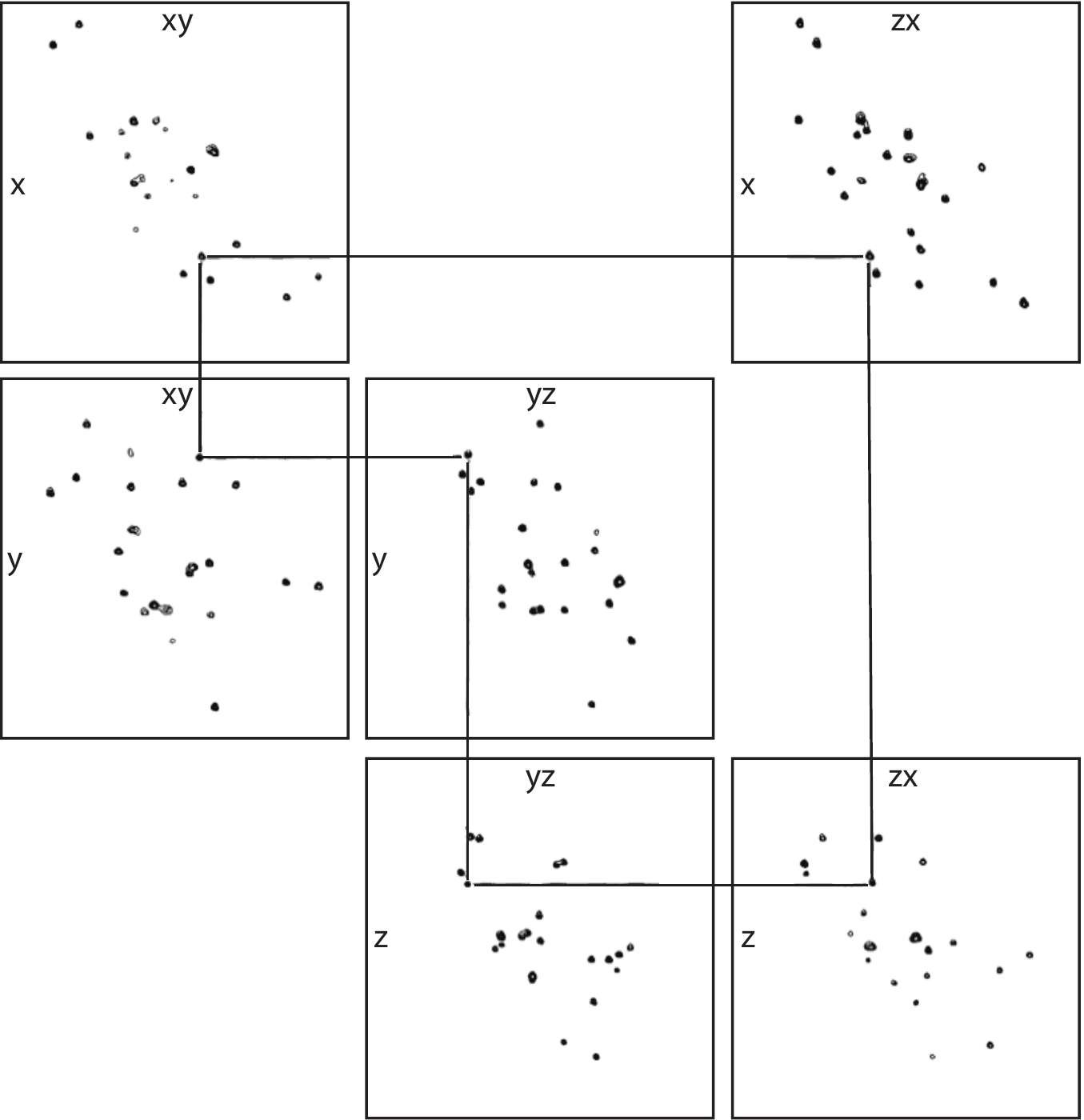}
	\caption{Examples of 2D spectra that correlate different crystal orientations obtained using the probe shown in Fig.~\ref{fig:sherwood}. The present experiments are $^{13}$C CP NMR spectra of sucrose. Adapted \cite{Sherwood_Alderman_Grant_1989} with permission.}
	\label{fig:sherwood_shift}
\end{figure}

Grant and co-workers \cite{Sherwood_Alderman_Grant_1989} took the idea of reducing the number of crystal orientations to its minimum in their design of the angle-flipping probe that allowed to establish a set of two-dimensional NMR spectra establishing pairwise correlations of six different crystal orientations using the probe shown in Fig.~\ref{fig:sherwood}. The probe design allows to visit six different crystal orientations required to determine the six parameters describing the magnitude and orientation of a chemical shift tensor. Through acquiring different correlation spectra, the overall correlation pattern in Fig.~\ref{fig:sherwood_shift} was obtained. This kind of correlation patterns provided a simple way to correlate a large number of shift tensors in crowded spectra.

%%%%%%%%%%%%%%%%%%%%%%%%%%%%%%%%%%%%%%%%%
% Applications
%%%%%%%%%%%%%%%%%%%%%%%%%%%%%%%%%%%%%%%%%
\section{Applications of Single-Crystal NMR}\label{sec:applications}

It is beyond the scope of this paper to review all SC NMR studies. Instead, this section highights selected studies that demonstrate the capabilities of SC NMR in various application areas.

\subsection{Carbon-13 Chemical Shift Tensors}

\begin{figure*}[pos=t]
	\centering
		\includegraphics[width=0.9\linewidth]{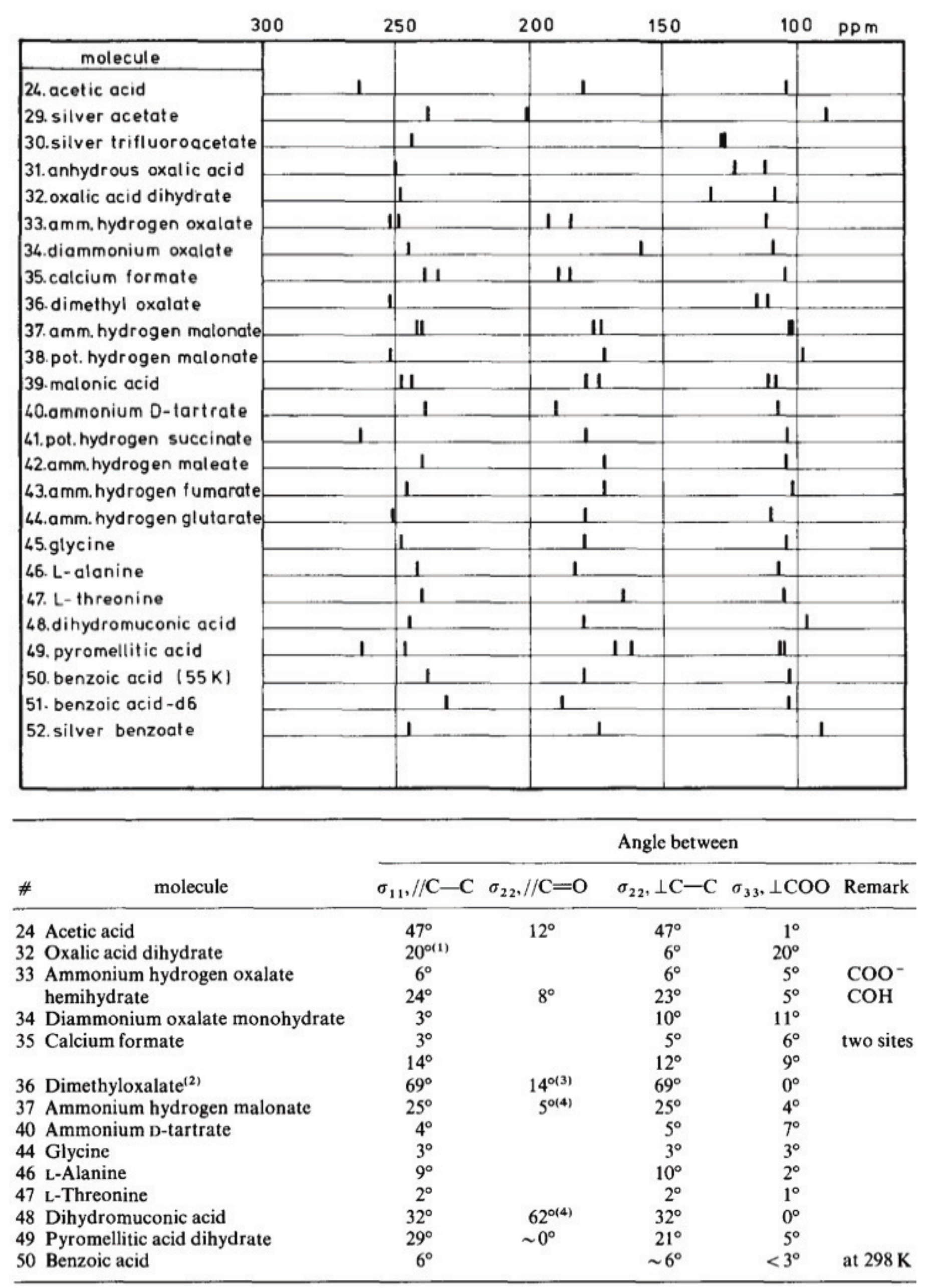}
	\caption{Summary of the magnitude (top) and orientation (bottom) of the principal elements of $^{13}$C chemical shift tensors in different compounds containing carboxylic groups. Adapted from  \cite{Veeman_1984} with permission.}
	\label{fig:veeman}
\end{figure*}

Just like a large number of liquid-state NMR studies have established empirical knowledge on $^1$H and $^{13}$C isotropic chemical shifts of different functional groups, a number of $^{13}$C SC NMR studies performed decades ago have provided valuable insight into the magnitudes and common orientations of the full chemical shift tensors for different chemical groups. Much of this work has been pioneered by Veeman \cite{van_dongen_1977, van_dongen_1978, torman_1978, Veeman_1981}, who also reviewed state-of-the-art of $^{13}$C chemical shift tensors in 1984 \cite{Veeman_1984}. Several later studies have added other types of functional groups, e.g. sugars \cite{Alderman_Sherwood_Grant_1993}. Reference \citealp{Veeman_1984} should be consulted for knowledge on the chemical shift tensors of both carboxylic, carbonylic, aromatic, and aliphatic groups. In Fig.~\ref{fig:veeman} we reproduce the findings only for carboxylic groups. General observations for different functional groups from the study of Veeman \cite{Veeman_1984} are summarized in the following.

For carboxyl groups, the $^{13}$C chemical shift principal  element with highest shift ($\delta_{zz} \approx$ 250 ppm) lies in the O-C-O plane and in-between the two C-O bonds, but the exact orientation depends on the protonation state of the carboxylic group. The principal element with lowest chemical shift ($\delta_{xx} \approx$ 100 ppm) is perpendicular to the O-C-O plane. The size of the last principal element lies approximately midway between the other principal elements: $\delta_{yy} \approx$ 175 ppm. These values are different for carbonyl groups, which have two principal elements around 250 ppm and for esters, where $\delta_{yy}$ decreases towards 130 ppm.

For aromatic carbons, the values of the three principal elements are $\delta_{zz} \approx$ 15 ppm, $\delta_{yy} \approx$ 150 ppm, and $\delta_{xx} \approx$ 225 ppm, with $\delta_{xx}$ approximately along the bond to the bound hydrogen or substituent and $\delta_{zz}$ perpendicular to the benzene ring.

The size and orientation of the $^{13}$C chemical shift tensors for methylene carbons vary much depending on the substituents, so it is difficult to obtain general trends for this. 

Methyl groups show relatively small shielding anisotropies with close to axial symmetry and the unique principal element aligned along the CH$_3$-C axis. The unique principal element has a value of $\delta_{zz} \approx$ 0 ppm, and the other elements are $\delta_{xx,yy} \approx$ 30 ppm. Interestingly, for methoxy groups, the unique element stays at a low value of $\delta_{zz} \approx$ 5 ppm, while the other elements increase to $\delta_{xx,yy} \approx$ 75 ppm.

\subsection{Biological Applications of Single-Crystal NMR}

\subsubsection{Nuclear Spin Interactions in Peptides}

\begin{figure}[pos=t]
	\centering
		\includegraphics[width=\linewidth]{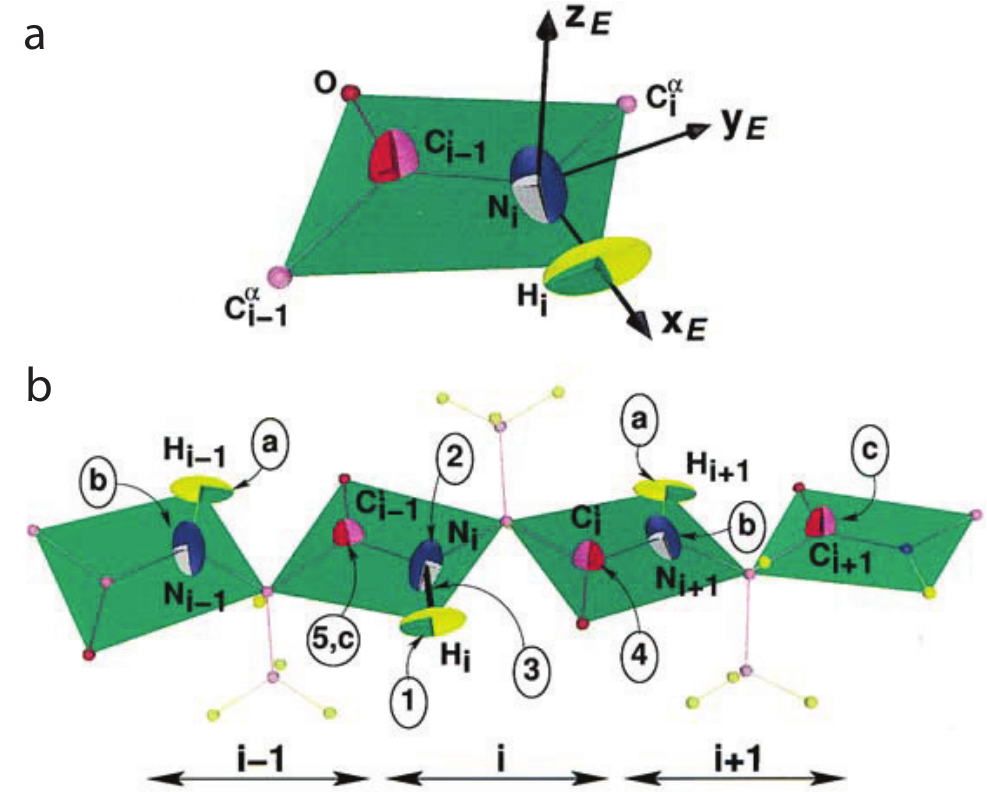}
	\caption{Representations generated by SIMMOL \cite{Bak_Schultz_Vosegaard_Nielsen_2002}  of a peptide plane (a) and peptide chain (b) showing typical orientations of the chemical shift tensors for the backbone atoms. Adapted from \cite{Vosegaard_Nielsen_2002} with permission.}
	\label{fig:peptide}
\end{figure}

Peptides and peptide mimics have been the target of several SC NMR studies. Particularly the orientation of the chemical shift tensors of the backbone atoms have been investigated, and it turns out that the orientations of the backbone chemical shift tensors of $^{15}$N for the amide nitrogen, $^1$H for the amide hydrogen, and $^{13}$C for the amide carbonyl atoms are largely defined by the geometry of the peptide plane as illustrated in Fig.~\ref{fig:peptide} \cite{Bak_Schultz_Vosegaard_Nielsen_2002} . The orientation of the chemical shift tensor for $^{13}$C in C$^\alpha$ have not been indicated in this figure, as its orientation depends strongly on the peptide torsion angles as investigated in detail by the pioneering work of Oldfield and co-workers \cite{sun_carbon13_2002}. The fact that the orientations of the backbone chemical shift tensors is mainly defined by the geometry of the peptide plane and less by the amino acid type forms the basis of oriented-sample solid-state NMR of proteins \cite{bechinger_flatcoil_1991, opella_structure_2004, Hansen_Bertelsen_Paaske_Nielsen_Vosegaard_2015}.

\begin{figure}[pos=t]
	\centering
		\includegraphics[width=0.6\linewidth]{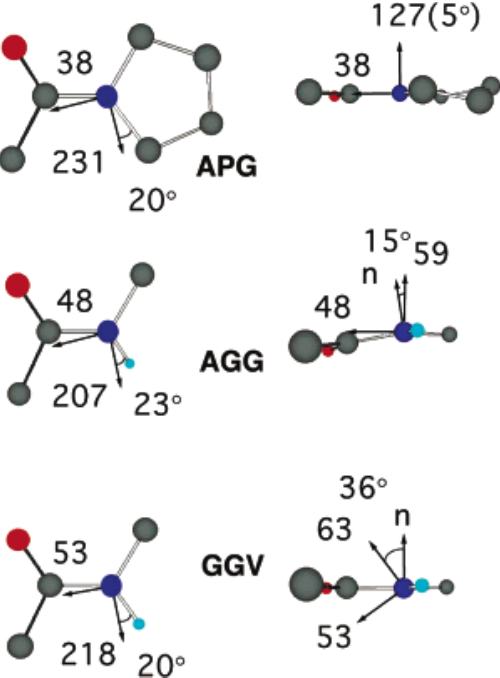}
	\caption{Orientations of $^{15}$N chemical shift tensors for tripeptides with $^{15}$N labelling of the middle amino acid. Views are along (left) and perpendicular to (right) the peptide plane normal ($n$). Reprinted from \cite{Waddell_Chekmenev_Wittebort_2005} with permission.}
	\label{fig:tripeptide}
\end{figure}

Harbison \etal~first determined a $^{15}$N chemical shift tensor in the dipeptide GlyGly \cite{Harbison_Jelinski_Stark_Torchia_Herzfeld_Griffin_1984}. More recently, Wadell \etal~studied a number of tripeptides \cite{Waddell_Chekmenev_Wittebort_2005}, for which the results are summarized in Fig.~\ref{fig:tripeptide}. Their findings were largely in agreement with previous determinations of the $^{15}$N chemical shift tensors determined by solid-state NMR of powdered samples \cite{teng_determination_1992, wu_simultaneous_1995, meng_tan_solidstate_1999}. Recently, $^{15}$N tensors in different peptides have been reviewed, both experimentally and numerically \cite{Harper_2017}. To summarize, the $^{15}$N chemical shift tensor has its low-field element ($\delta_{zz}\approx$ 220 ppm) arranged close to the N--H bond vector but deviating by ca 20$^\circ$ around the bilayer normal, while $\delta_{yy} \approx$ 120 ppm is almost parallel to the peptide-plane normal, although deviations by up to $36^\circ$ have been reported \cite{Waddell_Chekmenev_Wittebort_2005}, and it may be somewhat lower for glycine. The third element has a size of  $\delta_{xx} \approx$ 50 ppm.

\begin{figure}[pos=t]
	\centering
		\includegraphics[width=0.6\linewidth]{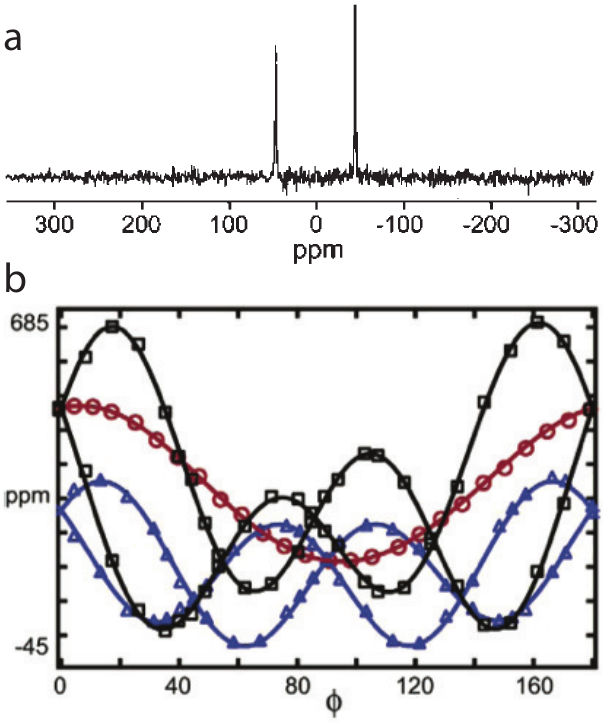}
	\caption{(a) Oxygen-17 NMR spectrum of a G[$^{17}$O]GV crystal showing the central transition. (b) Rotation plots for the resonances of the central transition of the same crystal for rotation around the crystallographic $a$ (black squares), $b$ (red circles), and $c^*$ (blue triangles) axes, respectively. The crystal belongs to the monoclinic space group $P2_1$ with two chemically equivalent but magnetically inequivalent sites. Reprinted from \cite{Waddell_Chekmenev_Wittebort_2006} with permission.}
	\label{fig:17o}
\end{figure}

Wittebort and co-workers \cite{Waddell_Chekmenev_Wittebort_2006, Chekmenev_Waddell_Hu_Gan_Wittebort_Cross_2006} presented a $^{17}$O SC NMR study of GGG, GGV, and AGG tripeptides with $^{17}$O enrichment of the carbonyl group of the middle amino acid. Selected results of this study are presented in Fig.~\ref{fig:17o} for GGV. Notably, the spectrum shown in Fig.~\ref{fig:17o} was obtained after $\sim$1k scans acquired in around one hour. The authors found that the most shielded principal element of the chemical shift tensor and the smallest electric-field gradient tensor element normal to the peptide plane, while the least shielded chemical shift element is in the peptide plane at an angle of $17^\circ$ away from the C=O bond and the largest electric-field gradient tensor element is in the peptide plane and perpendicular to the C=O bond.

\subsubsection{Phosphorous-31 Single-Crystal NMR Studies of Lipids and Lipid Mimics}

Phosphorous plays an important role in many areas of chemistry and biology. The $^{31}$P chemical shift has proven a sensitive reporter to the chemical environment of the phosphorous nucleus. Here, we will focus on lipids and lipid mimics, which are essential in solid-state NMR studies of membrane proteins. Some SC NMR studies in the 1970'es determined the chemical shift tensors of phosphates in lipid-like molecules \cite{Herzfeld_Griffin_Haberkorn_1978} and diphosphates  \cite{kohler_31p_1976}. The list of lipid-like molecules investigated by $^{31}$P SC NMR has grown since then \cite{Calsteren_Birnbaum_Smith_1987, Hauser_Radloff_Ernst_Sundell_Pascher_1988, Eichele_Wasylishen_2002}.

\begin{figure}[pos=t]
	\centering
		\includegraphics[width=0.6\linewidth]{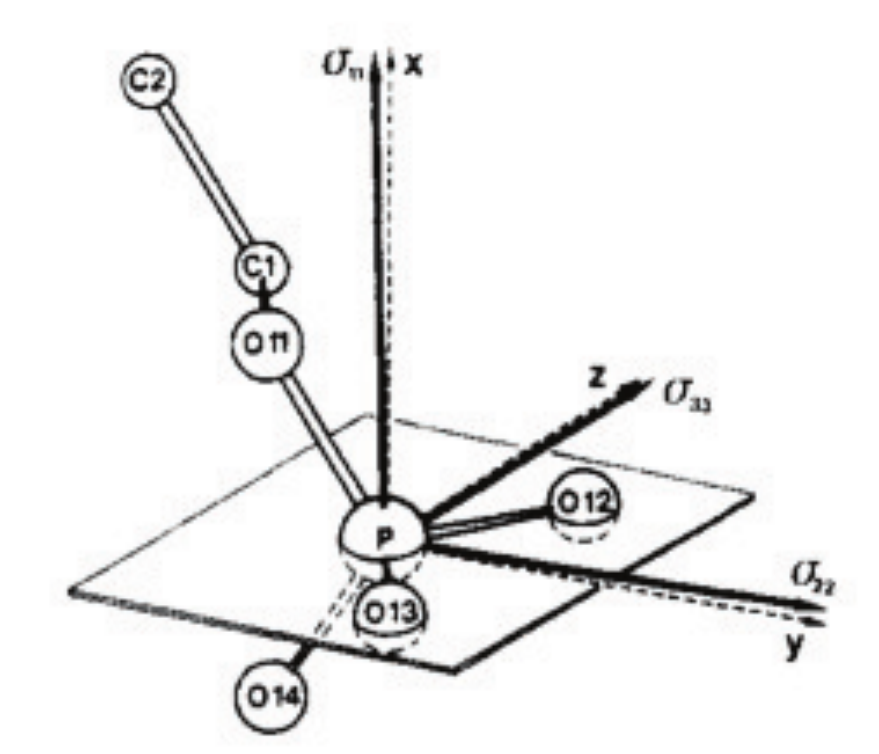}
	\caption{Orientation of the $^{31}$P chemical shift tensor relative to the phosphate group in the molecule l-hexadecyl-2-deoxyglycerophosphoric acid monohydrate. Reprinted from \cite{Hauser_Radloff_Ernst_Sundell_Pascher_1988} with permission.}
	\label{fig:31p}
\end{figure}

While the isotropic chemical shift may be used to report on variations in the chemical surroundings, these changes are small compared to the chemical shift anisotropy observed in solid samples. It turns out that the orientation of the chemical shift tensor is largely defined by the atoms in the first and second coordination sphere. The typical orientation of the principal elements of the chemical shift tensor  is shown in Fig.~\ref{fig:31p} using the nomenclature that phosphate oxygens O11 and O14 bind to the aliphatic chain and potential headgroup, respectively, while O12 and O13 do not bind any substituents \cite{Hauser_Radloff_Ernst_Sundell_Pascher_1988}. One principal element ($\delta_{yy} \approx -2$ ppm) intersects the P-O12 and P-O13 bonds and one principal element ($\delta_{xx}\approx -53$ ppm) is perpendicular to the plane spanned by O12, P, and O13. The last principal element ($\delta_{zz} \approx 58$ ppm) is also in this plane \cite{Herzfeld_Griffin_Haberkorn_1978, Calsteren_Birnbaum_Smith_1987, Hauser_Radloff_Ernst_Sundell_Pascher_1988}.

Knowledge of the $^{31}$P chemical shift tensor orientation in lipids has been important for several recent studies, for example when studying the dynamic averaging of the chemical shift tensor by combining NMR and molecular dynamics (MD) simulations \cite{Hansen_Vestergaard_Thogersen_Schiott_Nielsen_Vosegaard_2014, Vestergaard_Kraft_Vosegaard_Thogersen_Schiott_2015, weber_membrane_2016}. Commonly for these studies, the $^{31}$P chemical shift tensor was placed on the phosphorous atoms as described above for each snapshot of the MD simulation, and the average was calculated. Drobny and co-workers used the knowledge on the $^{31}$P chemical shift tensor in their study of binding of statherin to hydroxyapatite surfaces \cite{raghunathan_homonuclear_2006}.

\subsection{Materials Science Applications of Single-Crystal NMR}

This section presents SC NMR studies that have contributed to understanding of the nuclear spin interactions associated with isotopes relevant to materials science.

\subsubsection{Lithium-7 Single-Crystal NMR}

\begin{figure*}[pos=b]
	\centering
		\includegraphics[width=0.8\linewidth]{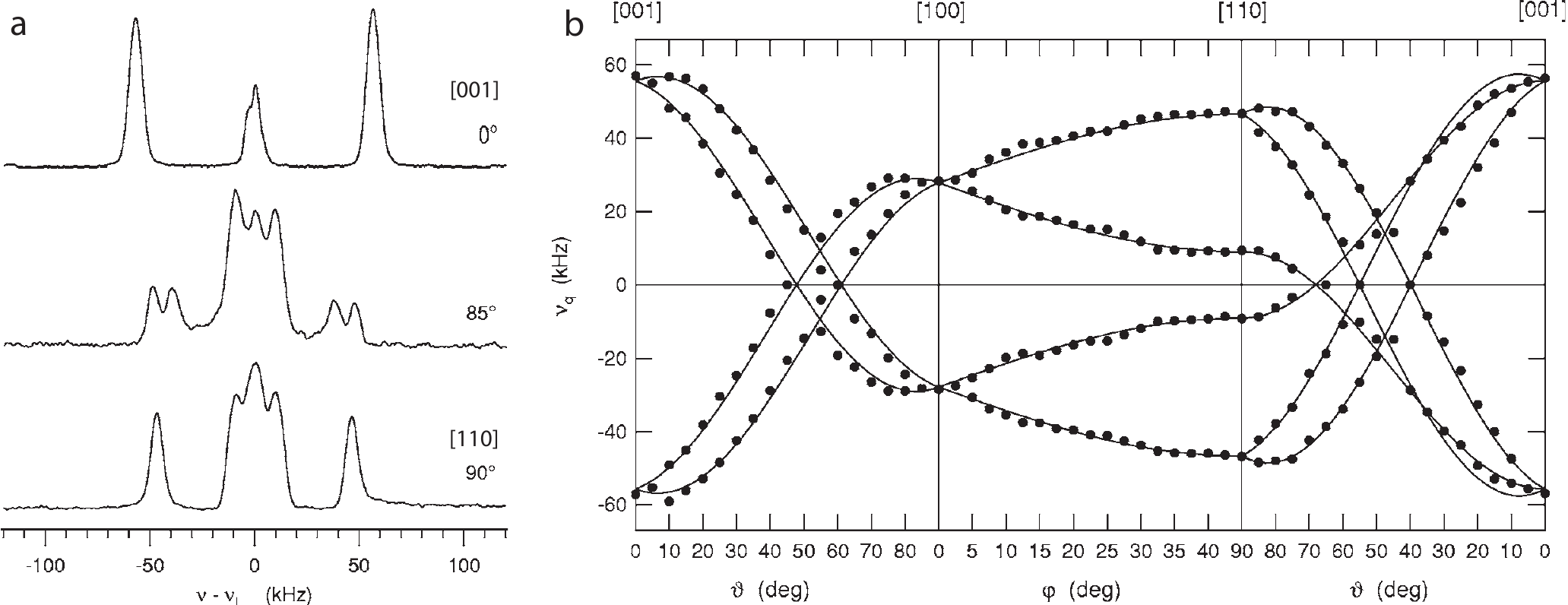}
	\caption{(a) $^7$Li SC NMR spectra of $\gamma$-LiAlO$_2$ for different orientations between the [001] and [110] axes. (b) Rotation plots for the $^7$Li resonances in  $\gamma$-LiAlO$_2$ for crystal rotations around the [010] (left), [001] (middle), and [$\overline{1}$10] (right) axes. Adapted from \cite{Indris_Heitjans_Uecker_Bredow_2006} with permission.}
	\label{fig:7li}
\end{figure*}

Lithium-7 NMR has many applications in materials science. One example are lithium aluminates, which are used in semiconductors and lithium electrodes and electrolytes. The chemical shift and quadrupole coupling parameters for different forms of LiAlO$_2$ have been investigated in an early $^7$Li NMR study of powdered samples \cite{muller_chemical_1983}. $\gamma$-LiAlO$_2$ was investigated by $^7$Li SC NMR and quantum-chemical calculations to investigate its local electronic structure \cite{Indris_Heitjans_Uecker_Bredow_2006}. Figure \ref{fig:7li} summarizes these results with selected $^7$Li SC NMR spectra and corresponding rotation plots and fitted resonance frequencies for different crystal orientations.

The structure of $\gamma$-LiAlO$_2$ consists of Li-O$_4$ tetrahedra sharing two oxygens with Al-O$_4$ tetrahedra. The largest principal element of the electric field gradient tensor at the position of the Li atoms is almost aligned with the crystallographic $c$ axis ($\sim 8^\circ$ deviation) and results in a $^7$Li quadrupole coupling of $C_\mathrm{Q} = 115.1$ kHz \cite{Indris_Heitjans_Uecker_Bredow_2006}.

\subsubsection{Aluminium-27 Single-Crystal NMR}

Aluminium-27 solid-state NMR has had a long tradition in materials science, since it provides valuable information about local conformation of Al in various materials. For example it is easy to distinguish Al$_\mathrm{IV}$, Al$_\mathrm{V}$, and Al$_\mathrm{VI}$ coordinations by their chemical shift \cite{haouas_recent_2016}. Because SC NMR allows to determine the tensorial nuclear spin interaction tensors with high precision, it has been used in studies of several crystalline materials \cite{filsinger_search_1997, Vosegaard_Jakobsen_1997, Vosegaard_Byriel_Pawlak_Wozniak_Jakobsen_1998, Bryant_1999, brauniger_local_2016, zeman_local_2020}. Due to this precision, \al\ SC NMR was used to observe chemical shift anisotropies of aluminum \cite{Vosegaard_Jakobsen_1997, Bryant_1999, zeman_local_2020}. In powdered samples, the central transition for \al\ is typically dominated by the second-order quadrupolar lineshape. In spite of this, Samoson \etal~\cite{samoson_chemical_1993} combined several experiments to determine the \al\ CSA from a powder sample of AlPO$_4$-21. Schurko \etal~\cite{schurko_characterization_1998} later found that the $^{27}$Al powder spectrum of AlCl$_3\cdot$OPCl$_3$ contains a clear signature of $^{27}$Al CSA.

\begin{figure}[pos=t]
	\centering
		\includegraphics[width=0.6\linewidth]{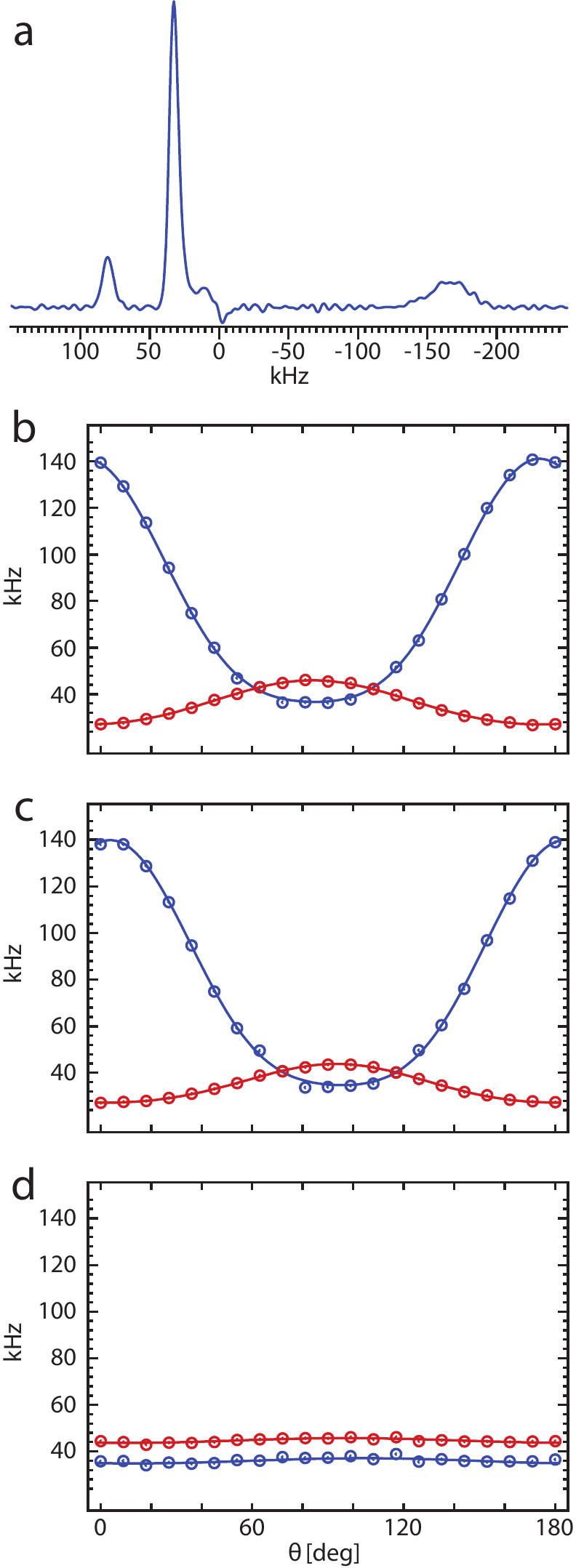}
	\caption{(a) Typical $^{27}$Al SC NMR spectrum of the alloy Al$_3$Ti showing two central transitions at 81 kHz and 33 kHz and a somewhat broader satellite transition at -165 kHz. (b-d) Rotation plots for the x (b), y (c), and z (d) axis, respectively \cite{vosegaard_al3ti}, showing data for site 1 in blue and site 2 in red.}
	\label{fig:al3ti}
\end{figure}

Aluminium-27 Knight shifts were reported by Bastow and Smith \cite{bastow_observation_1995}, who soon after also observed strong $^{27}$Al Knight shift anisotropies in aluminum-tin alloys \cite{smith_detection_1996}. Later, other Knight shift anisotropies have also been reported \cite{yuan_solid_2012}. We investigated the alloy Al$_3$Ti by $^{27}$Al SC NMR as summarized in Fig.~\ref{fig:al3ti} \cite{vosegaard_al3ti} and found that the two sites located on four-fold symmetry axes rendering their nuclear spin interactions axially symmetric. The quadrupole couplings are $C_\mathrm{Q} = 11.10 \pm 0.18$ MHz (site 1) and $C_\mathrm{Q} = 3.03 \pm 0.03$ MHz (site 2) and with strong Knight shift anisotropies of $K_\mathrm{aniso} = -491 \pm 3$ ppm (site 1) and $K_\mathrm{aniso} = 72 \pm 2$ ppm (site 2).

\subsubsection{Vanadium-51 Single-Crystal NMR}

Man and co-workers \cite{man_application_1994} used nutation experiments on single crystals of BiVO$_4$ to determine the quadrupole coupling for \vv\ in such crystals by fitting the nutation frequencies for the different transitions, by exploiting the fact that these are sensitive to the first-order quadrupole coupling interaction that mixes with the RF pulses.

\begin{figure}[pos=t]
	\centering
		\includegraphics[width=\linewidth]{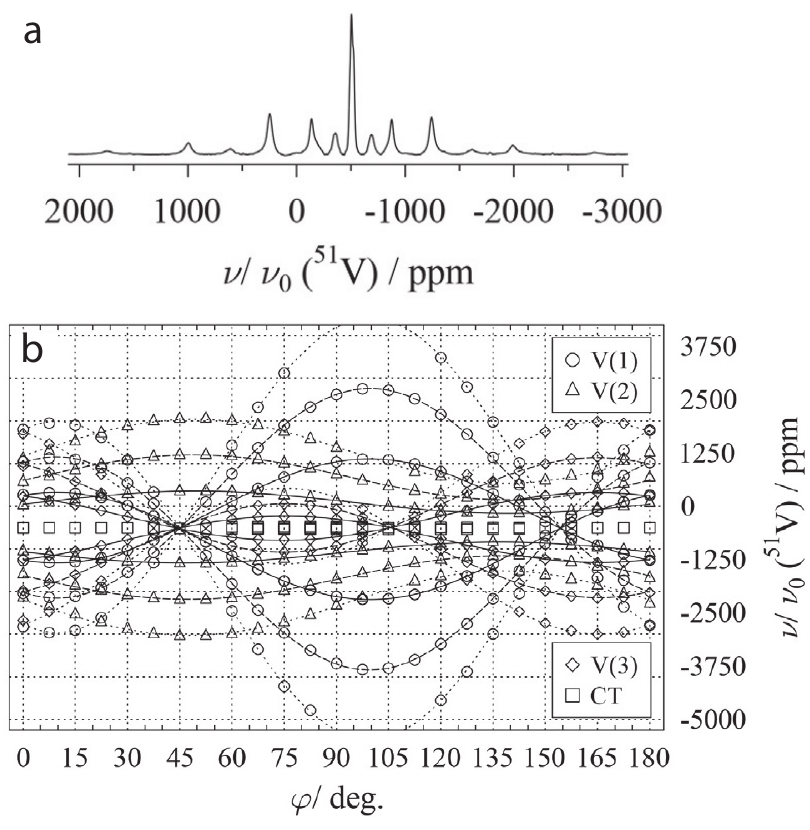}
	\caption{(a) Vanadium-51 NMR spectra for a single crystal of Pb$_5$(VO$_4$)$_3$Cl at different orientations relative to the magnetic field. (b) Rotation plot for the five magnetically inequivalent \vv\ sites in the hexagonal crystal structure. Adapted from \cite{Zeman_Hoch_Hochleitner_Bruniger_2018} with permission.}
	\label{fig:51v}
\end{figure}

Zeman \etal~\cite{Zeman_Hoch_Hochleitner_Bruniger_2018} characterized the vanadinite mineral Pb$_5$(VO$_4$)$_3$Cl by \vv\ and $^{207}$Pb SC NMR. Due to the location of lead and vanadium atoms at different special positions in the hexagonal crystal structure, it was possible to use the symmetries of the different magnetically inequivalent sites to determine the full nuclear spin interaction tensors using only a single rotation axis as discussed in section \ref{sec:one} above. For illustration, Fig.~\ref{fig:51v} shows typical \vv\ SC NMR spectra and the rotation plot for all six satellite transitions (spin $I=7/2$).

\subsubsection{Manganese-55 Single-Crystal NMR}

\begin{figure}[pos=t]
	\centering
		\includegraphics[width=0.9\linewidth]{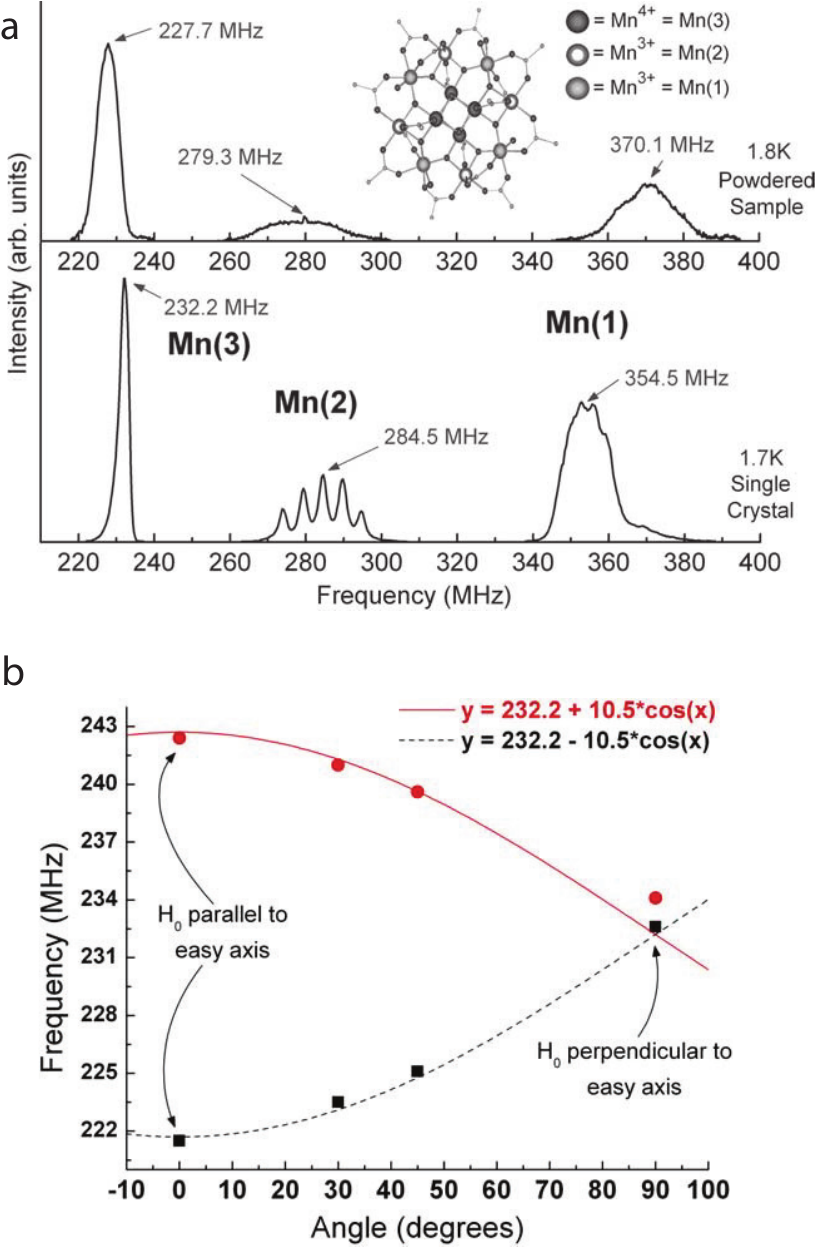}
	\caption{(a) Manganese-55 powder (top) and single-crystal (bottom) NMR spectra of Mn$_{12}$-BrAc at zero external magnetic field. (b) Orientational dependence of the peak splitting observed in a single crystal experiencing an external magnetic field of 1 T. Adapted from \cite{Harter_Chakov_Achey_Reyes_Kuhns_Christou_Dalal_2005} with permission.}
	\label{fig:55Mn}
\end{figure}

Manganese-55 NMR is commonly used in the study of materials. Here, in an interesting but unconventional applications of SC NMR, Harter \etal\ \cite{Harter_Chakov_Achey_Reyes_Kuhns_Christou_Dalal_2005, Harter_Chakov_Roberts_Achey_Reyes_Kuhns_Christou_Dalal_2005} used \mn\ SC NMR to study the single-molecule magnet [Mn$_{12}$O$_{12}$-(CH$_2$BrCOO)$_{16}$4(H$_2$O)] $\cdot$4CH$_2$Cl$_2$ (Mn$_{12}$-BrAc). Figure \ref{fig:55Mn}a shows the \mn\ NMR spectra of a magnetically oriented powder sample of Mn$_{12}$-BrAc (top) and a single crystal (bottom) at zero external magnetic field. This figure demonstrates how the three manganese sites experience different magnetic fields produced by the molecules. In the single-crystal case, the three sites experience magnetic field strengths of 22.11 T (Mn(3)), 27.10 T (Mn(2)), and 33.76 T (Mn(1)). The difference observed compared to the powder sample, where the single-molecule magnets tend to co-align, was ascribed to temperature and pressure changes arising when crushing the sample.

Further analyses of the single-molecule magnets were made by investigating the splitting of the Mn(3) resonance for different orientations of crystal relative to an external applied magnetic field (1 T). These results are reported in Fig.~\ref{fig:55Mn}b and discussed further in the original paper \cite{Harter_Chakov_Achey_Reyes_Kuhns_Christou_Dalal_2005}.

\subsubsection{Zinc-67 Single-Crystal NMR}

\begin{figure}[pos=t]
	\centering
		\includegraphics[width=\linewidth]{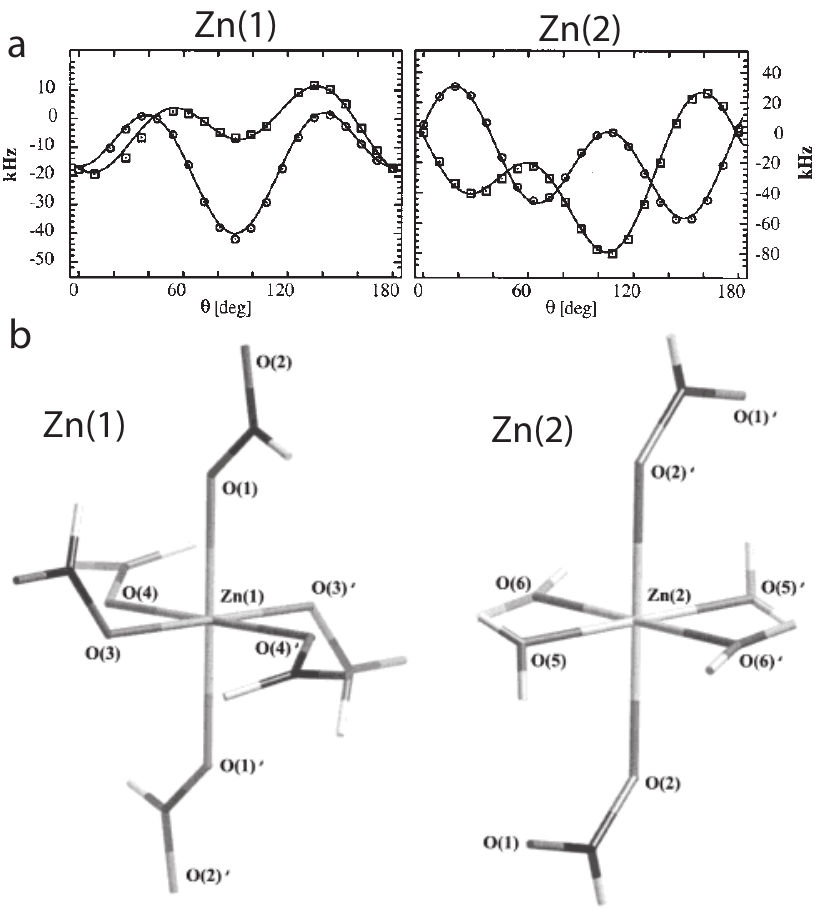}
	\caption{(a) Rotation plots for the z axis for the two $^{67}$Zn sites in a crystal of zink formate dihydrate. (b) Structures of the anhydrous site, Zn(1), and hydrated site, Zn(2) in zink formate dihydrate. Adapted from \cite{lipton_67zn_2002} with permission.}
	\label{fig:67zn}
\end{figure}

The $^{67}$Zn SC NMR study of zink formate dihydrate \cite{lipton_67zn_2002} showed peaks for the two octahedral Zn sites in this crystal, corresponding to an anhydrous, Zn(1), coordinated by six formate groups and a site with two formates and four water molecules in the first coordination sphere, Zn(2). This fact was used for assignment of the two sites; $^1$H-$^{67}$Zn Cross-polarization experiments with varying contact times and for samples crystallized from H$_2$O and D$_2$O identified the site in close proximity to water. Figure \ref{fig:67zn} shows the local structures of the zink atoms and selected results of this study. The $^{67}$Zn SC NMR spectra allowed extraction of $^{67}$Zn quadrupole coupling parameters but was not sufficiently sensitive to measure the $^{67}$Zn chemical shift anisotropy. The orientation of the electric-field gradient tensors was such that for the hydrated Zn(2) site, $V_{zz}$ was along the axis of the two formyl groups. For the other site, the $V_{zz}$ element was aligned $\sim 16^\circ$ away from the shortest Zn--O(4) bond.

We carried out a $^{67}$Zn study of zink acetate dihydrate \cite{Vosegaard_Andersen_Jakobsen_1999}, for which we did observe a non-neglectible chemical shift anisotropy. The effect of the CSA was small; When neglecting the CSA and when including a CSA of $\delta_\mathrm{aniso} = 33$ ppm, the resulting quadrupole parameters deviated by only 3 \%, but the fit between experiments and simulations was best when including the CSA parameters. In zink acetate dihydrate the coordination of zink is different than in zink formate dihydrate, as the zink site is coordinated by two water molecules and four oxygens from two acetate ions. We found that the chemical-shift and EFG tensors were coaligned within experimental precision.

\subsubsection{Gallium-69 and -71 Single-Crystal NMR}

Gallium solid-state NMR of both the $^{69}$Ga (spin $I=3/2$, 60.4\% natural abundance, quadrupole moment $0.168\cdot 10^{-24}$ e cm$^2$) and $^{71}$Ga (spin $I=3/2$, 39.6\% natural abundance, quadrupole moment $0.106\cdot 10^{-24}$ e cm$^2$) isotopes is challenging to perform due to their large quadrupole moments implying that even the second-order quadrupolar lineshape may be up to several hundreds of kHz broad \cite{Massiot_Vosegaard_Magneron_Trumeau_Montouillout_Berthet_Loiseau_Bujoli_1999}. Hence, for powdered samples it has been necessary to use advanced techniques such as dynamic-angle spinning to achieve site-specific resolution \cite{massiot_6971_1995}. We later proposed a projection-reconstruction method that used one-dimensional spectra from different magnetic field strengths to reconstruct a two-dimen\-sional correlation experiment with spectral resolution \cite{Vosegaard_Massiot_2007}. Most interestingly, Hung and Gan recently demonstrated that by using low-power STMAS it is now possible to resolve such sites, which seems to be of broad applicability \cite{hung_lowpower_2020}.

Inspired by the extensive work of Massiot and others on $^{69/71}$Ga solid-state NMR (reviewed in ref.~\citealp{Massiot_Vosegaard_Magneron_Trumeau_Montouillout_Berthet_Loiseau_Bujoli_1999}), we were interested in establishing the chemical shift and quadrupole coupling tensors for these isotopes in different compounds \cite{Vosegaard_Massiot_Gautier_Jakobsen_1997, Vosegaard_Byriel_Binet_Massiot_Jakobsen_1998}.

\begin{figure}[pos=t]
	\centering
		\includegraphics[width=\linewidth]{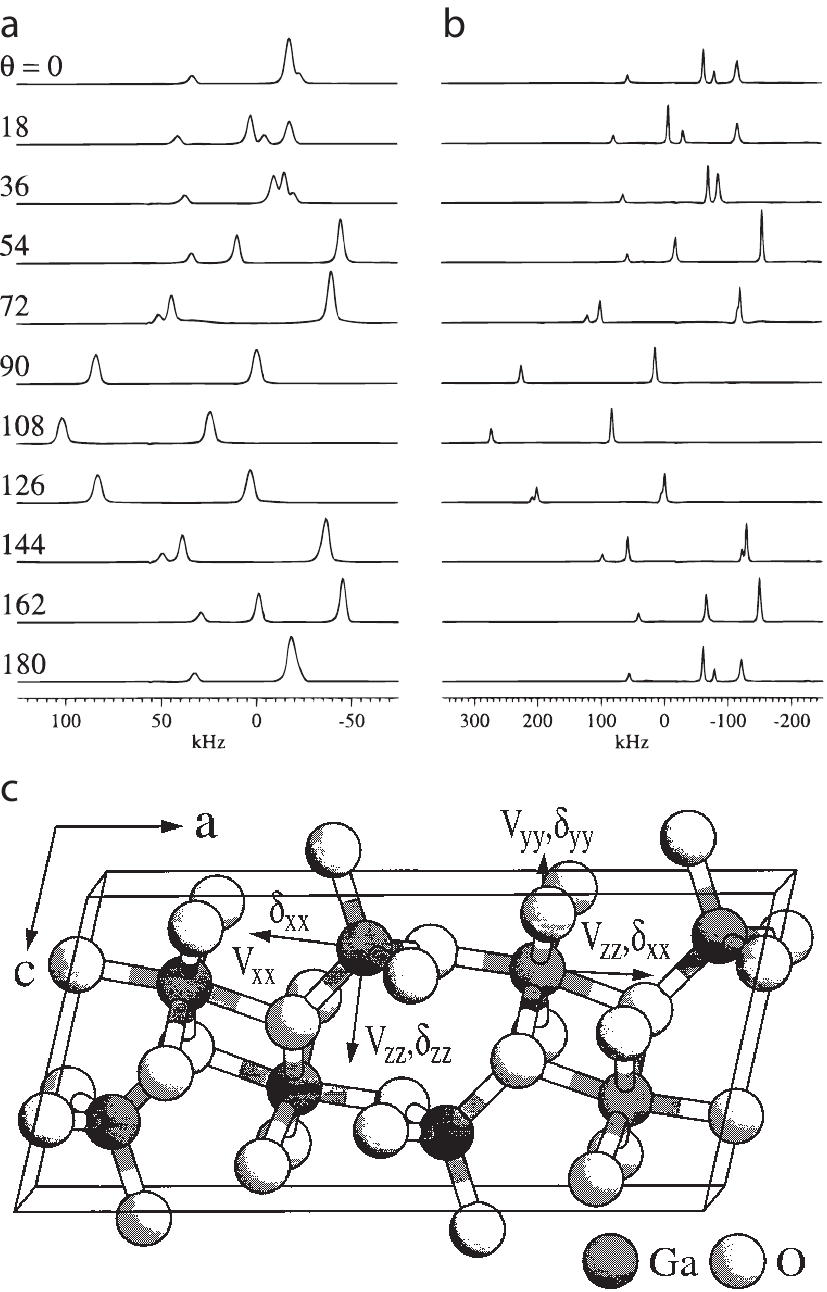}
	\caption{Gallium-71 (a) and Gallium-69 (b) SC NMR spectra for a crystal of $\beta$-Ga$_2$O$_3$ for rotation around the $x$ axis. Due to twinning, both a strong and a weak set of peaks is observed for both isotopes. (c) Orientation of the chemical shift and electric-field gradient tensors in the crystal frame. Adapted from \cite{Vosegaard_Byriel_Binet_Massiot_Jakobsen_1998} with permission.}
	\label{fig:Ga}
\end{figure}

Our $^{69/71}$Ga SC NMR study of $\beta$-Ga$_2$O$_3$ is summarized in Fig.~\ref{fig:Ga}. Both the $^{71}$Ga and $^{69}$Ga SC NMR spectra show the expected site-specific resolution and allowed to determine the very large quadrupole coupling and chemical shift with good precision \cite{Vosegaard_Byriel_Binet_Massiot_Jakobsen_1998}.

\subsubsection{Cadmium-113 Single-Crystal NMR}

In the 1980'es and 1990'es, \cd\ solid-state NMR obtained considerable interest, as \cd\ is a spin-1/2 nucleus NMR nucleus with a high gyromagnetic ratio (comparable to $^{13}$C), a reasonable natural abundance (12.2 \%), and a large chemical shift dispersion making it a good NMR nucleus. Adding to this, it is interesting that Cd may replace zinc and calcium in various biological contexts \cite{ellis_cadmium113_1983}. 

To establish the knowledge on how different coordinations and functional groups attached to cadmium affect the chemical shift, a large number of \cd\ SC NMR studies were performed to determine the magnitude and orientation of \cd\ CSA tensors in different molecules 
\cite{Honkonen_Doty_Ellis_1983, Honkonen2, Ganapathy_Chacko_Bryant_1984, Honkonen3, kennedy5, Kennedy_Ellis_Jakobsen_1990, rivera_cadmium113_1990}.

\begin{figure}[pos=t]
	\centering
		\includegraphics[width=0.8\linewidth]{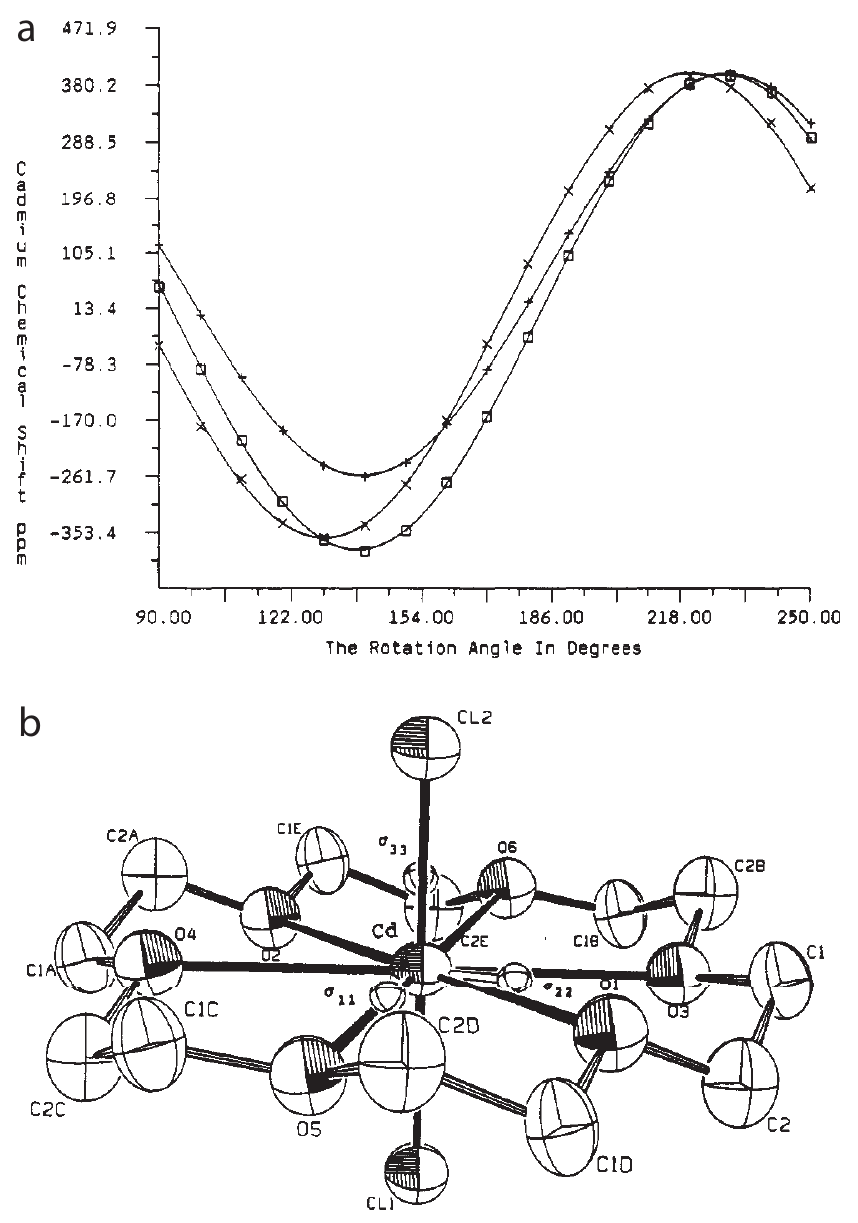}
	\caption{(a) Cadmium-113 chemical shifts in a single crystal of CdCl$_2\cdot$18-crown-6 for different orientations resulting from rotation of the crystal approximately around the crystallographic $a$ ($\times$), $b$ ($\square$), and $c$ ($+$) axis. (b) Ortep-representation of the molecular structure of CdCl$_2\cdot$18-crown-6 with indication of the orientation of the chemical shift tensor. Reproduced from \cite{Kennedy_Ellis_Jakobsen_1990} with permission.}
	\label{fig:113Cd}
\end{figure}

Figure \ref{fig:113Cd} summarizes the \cd\ SC NMR study of CdCl$_2\cdot$18-crown-6 \cite{Kennedy_Ellis_Jakobsen_1990}. This crown ether crystallizes in the trigonal space group $R3$ with one molecule in the unit cell. The rotation plots resulting from rotation approximately around the crystallographic axes are shown in Fig.~\ref{fig:113Cd}a. Analyses of these rotation plots resulted in chemical shift principal elements of $\delta_{xx}$ = 414 ppm, $\delta_{yy}$ = 385 ppm, and $\delta_{zz}$ = $-700$ ppm. The unique principal element aligned perpendicular to the flat crown-ether ring and exhibits an unusually large shielding \cite{Kennedy_Ellis_Jakobsen_1990}.

\subsubsection{Lead-207 Single-Crystal NMR}

\begin{figure}[pos=t]
	\centering
		\includegraphics[width=\linewidth]{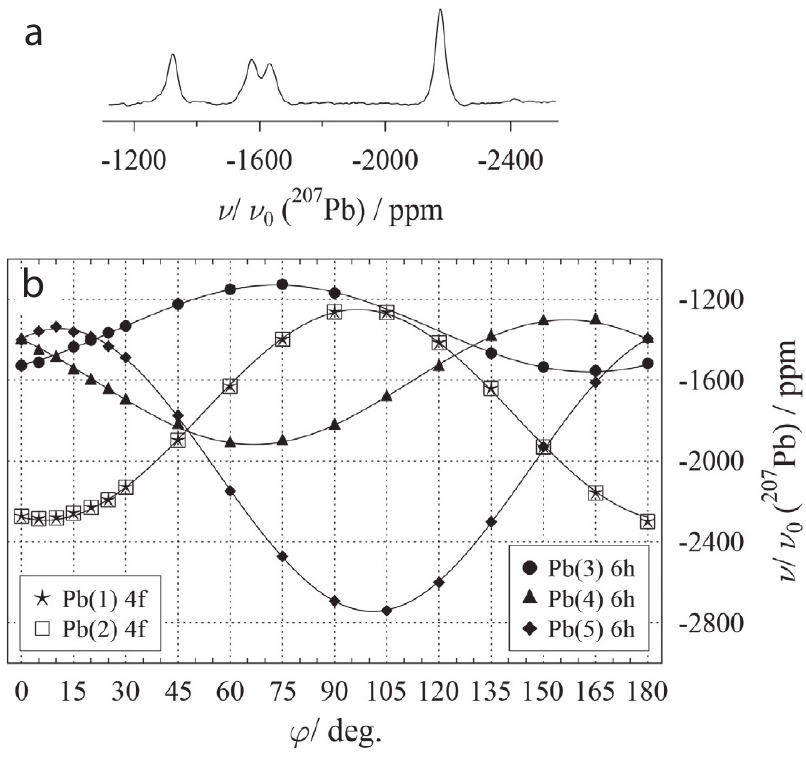}
	\caption{(a) Lead-207 NMR spectra for a single crystal of Pb$_5$(VO$_4$)$_3$Cl at different orientations relative to the magnetic field. (b) Rotation plot for the four resolved \pb\ sites in the hexagonal crystal structure. Adapted from \cite{Zeman_Hoch_Hochleitner_Bruniger_2018} with permission.}
	\label{fig:207Pb}
\end{figure}

Although \pb\ NMR has been extensively used to study different lead-containing solids \cite{dybowski_solid_2002}, to our knowledge there have only been few \pb\ SC NMR studies in the past 
\cite{luders_nuclear_1972, laguta_investigation_1994}, until this has recently been brought up by Bräuniger and co-workers \cite{Zeman_Moudrakovski_Hoch_Hochleitner_Schmahl_Karaghiosoff_Bruniger_2017, Zeman_Hoch_Hochleitner_Bruniger_2018, zeman_singlecrystal_2019, zeman_determination_2019}. In their combined \vv\ and \pb\ SC NMR study of Pb$_5$(VO$_4$)$_3$Cl \cite{Zeman_Hoch_Hochleitner_Bruniger_2018}, they observed the expected \pb\ resonances originating from three magnetically inequivalent  lead atoms in pos 6h and two in pos 4f. The latter two are not resolved, as they are both located on a three-fold symmetry axis aligned along the $c$ axis. Figure \ref{fig:207Pb} reports a typical \pb\ SC NMR spectrum of Pb$_5$(VO$_4$)$_3$Cl as well as the rotation plots for the four resolved sites. The chemical shift parameters were obtained by rotation around a single rotation axis and agreed well with previous data, while comparison with DFT did not provide good agreement, probably because the DFT algorithms used were not fully adapted to heavy atoms like \pb\ \cite{Zeman_Hoch_Hochleitner_Bruniger_2018}.

\subsection{Physical Properties Characterized by Single-Crystal NMR}

Single-crystal NMR is typically used to exploit the nuclear spin interactions of particular nuclei as reviewed in the previous section. Here, we focus on applications of SC NMR to study physical properties of crystals like phase transitions and twinning.

\subsubsection{Phase transitions}

The study of phase transitions by SC NMR is not so obvious, at least one should require that the crystal undergoes a single crystal to single crystal transformation, in which case SC NMR may provide a higher level of detail than obtainable from powdered samples.

\begin{figure}[pos=t]
	\centering
		\includegraphics[width=0.6\linewidth]{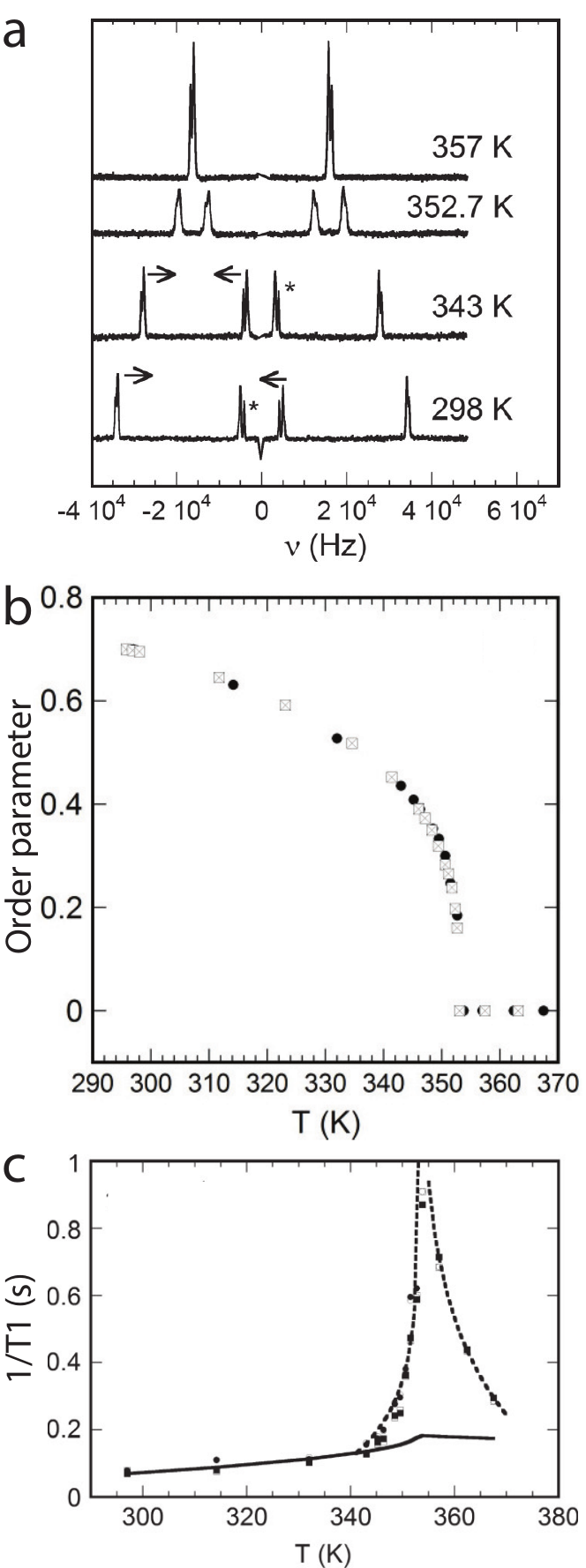}
	\caption{(a) Deuterium SC NMR spectra of KDCO$_3$ at different temperatures. (b) Order parameters measured from quadrupolar parameters for $^2$H (\Circle) and $^{39}$K ($\boxtimes$) and (c) inverse $^2$H spin-lattice relaxation time (1/T1) in KDCO$_3$ as a function of temperature. Adapted from \cite{Odin_2008} with permission.}
	\label{fig:Odin}
\end{figure}

Odin \cite{Odin_2008} studied a phase transition in KHCO$_3$, where the carbonates form dimers that are hydrogen-bonded in two different ways. At low temperature, the two configurations result in distinct resonances in $^2$H and $^{39}$K SC NMR (see typical $^2$H SC NMR spectra in Fig.~\ref{fig:Odin}a), whereas the peaks collapse at higher temperatures above the phase transition.

The first-order quadrupolar splitting ($^2$H) and second-order quadrupolar shift ($^{39}$K) was converted into an order parameter, which is depicted in Fig.~\ref{fig:Odin}b, clearly illustrating the sensitivity of the used NMR measurements to observe the phase transition.

A final measurement in the work of Odin \cite{Odin_2008} was the measurement of the $^2$H spin-lattice relaxation time T1 as a function of temperature as reproduced in Fig.~\ref{fig:Odin}c. This measurement also nicely shows the critical response of the spin system around the phase transition temperature.

\subsubsection{Relaxation}

Following the last results in the previous section, it is relevant to report other studies of the relaxation behaviour of spin systems in single crystals.

Skibsted and co-workers \cite{Andersen_Jakobsen_Skibsted_2005} assessed the temperature dependence of the transverse relaxation time for \al\ in in a single crystal of alun, KAl(SO$_4$)$_2\cdot$12H$_2$O, through measurement of the linewidths of the different transitions for \al. 

\begin{figure}[pos=t]
	\centering
		\includegraphics[width=\linewidth]{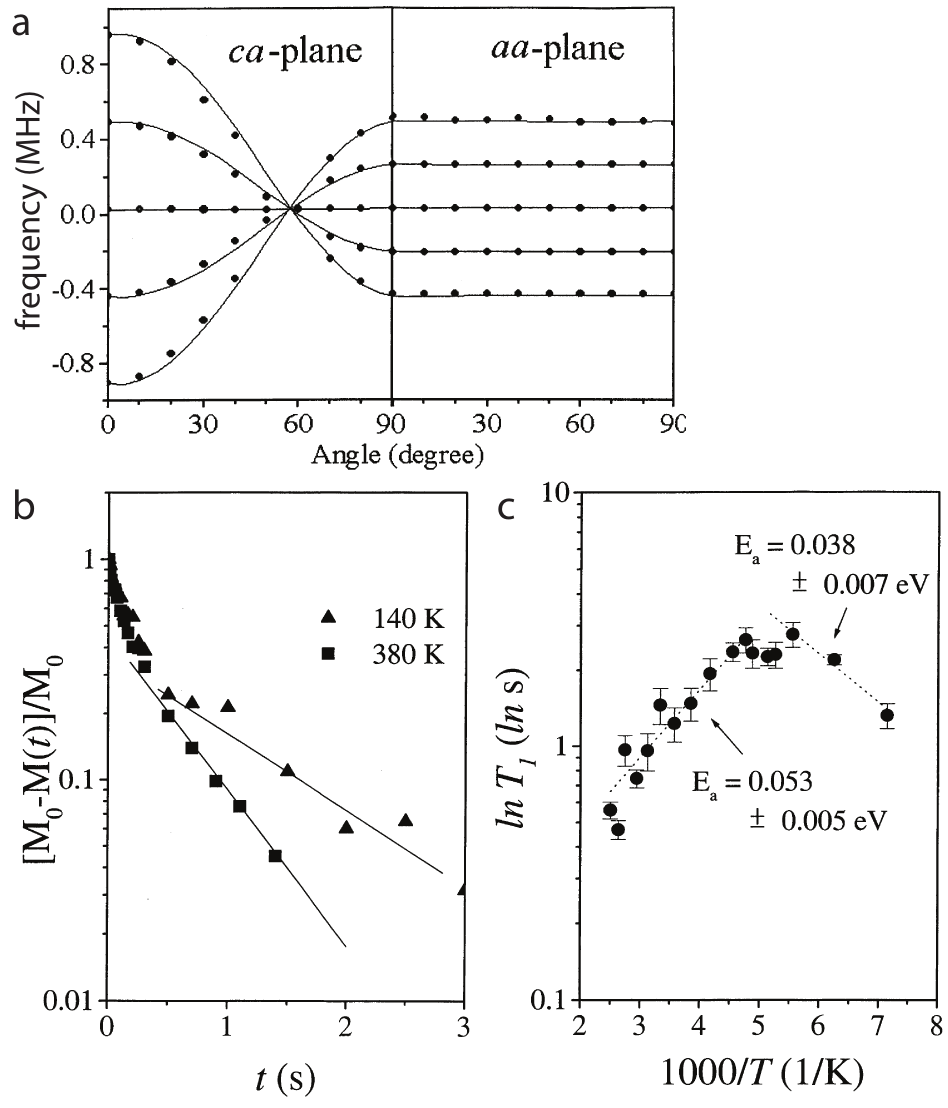}
	\caption{(a) Rotation plots for the \al\ resonances in a single crystal of emerald rotated as indicated in the figure. (b) \al\ magnetization build-up curves at different temperatures using saturation-recovery experiments. (c) Temperature dependence of the spin-lattice relaxation time T1 for \al\ in emerald. Adapted from \cite{kim_27al_2000} with permission.}
	\label{fig:t1}
\end{figure}

Kim \etal\ \cite{kim_27al_2000} studied the spin-lattice relaxation times in a single crystal of emerald, Be$_3$Al$_2$Si$_6$O$_{18}$:Cr$^{3+}$, as a function of temperature as reviewed in Fig.~\ref{fig:t1}. From the SC NMR data it was possible to determine the relaxation mechanisms being dominated by a two-phonon Raman relaxation process at high temperature, while paramagnetic relaxation caused by Cr$^{3+}$ impurities were dominant at lower temperatures.

\subsubsection{Crystal defects}

Verkhovskii \emph{et al.} \cite{Verkhovskii_1999, verkhovskii_isotopic_2003} investigated the disorder in single crystals caused by the presence of different isotopes by $^{73}$Ge SC NMR. Although this finding is somewhat surprising, their study of single crystals of germanium, seems convincing. Germanium was chosen as it is possible to grow nearly perfect cubic crystals of Ge without structural defects. The findings of this study was that it is possible to quantify the amount of crystal disorder induced by different isotopes through random noncubic crystal field interactions.

In our $^{69/71}$Ga SC NMR study of $\beta$-Ga$_2$O$_3$ \cite{Vosegaard_Byriel_Binet_Massiot_Jakobsen_1998} we did not manage to grow proper single crystals; all crystals were twinned. The crystal chosen for the SC NMR investigation displayed two sets of resonances with approximate intensity ratio 2:1 for the two twins (see Fig.~\ref{fig:Ga} for details). It turns out that twinning is common in $\beta$-Ga$_2$O$_3$ \cite{geller_crystal_1960, wolten_determination_1976}, and while twinning is typically problematic in XRD, in favourable cases it may be advantageous in SC NMR: We realised that the two sets of resonances led to tensors with relative orientations that could reveal the twin law, i.e. the common ways of defective crystal growth, responsible for the twinning of the $\beta$-Ga$_2$O$_3$ crystal \cite{Vosegaard_Byriel_Binet_Massiot_Jakobsen_1998}. This way, we obtained valuable information about the crystal morphology and the orientation of the crystallographic axes directly from the SC NMR data.

%%%%%%%%%%%%%%%%%%%%%%%%%%%%%%%%%%%%%%%%%
% ALTERNATIVE APPROACHES
%%%%%%%%%%%%%%%%%%%%%%%%%%%%%%%%%%%%%%%%%
\section{Alternative Approaches to Obtain Full Nuclear Spin Interaction Tensors}\label{sec:alternative}

There have been several attempts to obtain full nuclear spin interaction tensor information by other means than SC NMR. Measurement of dipolar couplings in combination with other tensorial interactions have been used in many cases. However, a few studies have pursued different approaches as reviewed in the following.

\subsection{Powder samples}

Powder samples contain a large number of small crystallites with random orientations. Recording NMR spectra of such static samples results in powder spectra, which have broad intensity distributions with shoulders and singularities. A few attempts have been made to spectroscopically filter such spectra to provide simpler spectra. Pines and co-workers \cite{Swiet_Tomaselli_Pines_1998} realised that the orientation-dependent recoupling strength of different rotary-resonance recoupling conditions may lead to dephasing of signals from all crystallites except for a few special combinations of the $\alpha$ and $\beta$ angles describing the orientation of the principal axis system. By repeating combinations of $N=1$ and $N=2$ rotary-resonance recoupling they demonstrated that it is possible to achieve NMR spectra of CSA patterns, where the powder pattern is replaced by three peaks at the frequencies of the principal values of the CSA tensor potentially scaled by the scaling factor of the pulse sequence \cite{Swiet_Tomaselli_Pines_1998}.

While the approach of Pines and co-workers did not provide information about the orientation of the chemical shift tensor, it demonstrated that it is possible to selectively observe specific crystallites (or rather crystallite orientations) in the sample. This approach has been further elaborated by Pell \etal\ \cite{Pell_Pintacuda_Emsley_2011}, who established a method to obtain information about the orientation of the tensorial nuclear spin interactions from spinning powders.

\begin{figure*}[pos=t]
	\centering
		\includegraphics[width=0.8\linewidth]{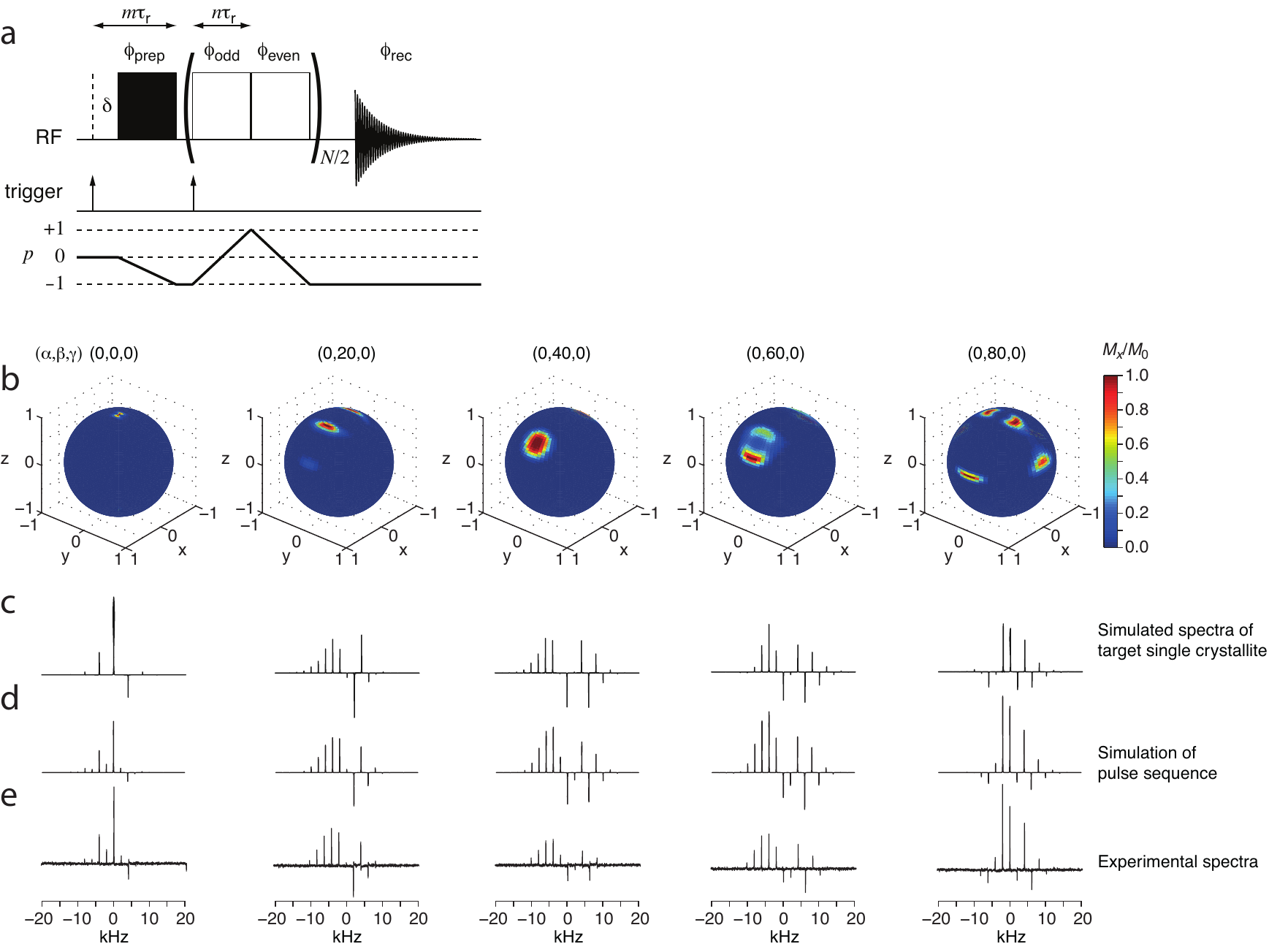}
	\caption{(a) Pulse sequence with a train of crystallite-selective pulses used for selective excitation of distinct crystallites. (b) Simulated excitation profiles targeting different crystallites indicated by ($\alpha, \beta, \gamma$). (c-e) Simulated (c,d) and experimental (e) spectra for the different excitation profiles. The simulations show the spectra for the specified crystallite (c) and result of the pulse sequence (d), respectively. Adapted from \cite{Pell_Pintacuda_Emsley_2011} with permission.}
	\label{fig:emsley}
\end{figure*}

Figure \ref{fig:emsley}a shows the pulse sequences used to generate the crystallite-selective excitation. As a result of this pulse sequence, only crystallites with a specific orientation relative to the rotor frame are excited. The ability to selectively excite different crystallite orientations was assessed by simulations as reproduced in Fig.~\ref{fig:emsley}b. The corresponding experiments and simulations for $^{13}$C MAS ($\nu_R = 2$ kHz) NMR of [1-$^{13}$C]-alanine are presented in Figs.~\ref{fig:emsley}c-\ref{fig:emsley}e. This method represents an interesting alternative to SC NMR studies, although issues like the orientational specificity and sensitivity needs be investigated further for practical uses of the method on real systems.

\subsection{Magnetic orientation}

Many samples display magnetic susceptibility anisotropy that allows for magnetic orientation of the samples. Model systems for biological cell membranes like bicelles align in strong magnetic fields allowing to obtain orientation-specific information about the spin system \cite{opella_structure_2004, Hansen_Bertelsen_Paaske_Nielsen_Vosegaard_2015}.

Likewise, the \mn\ NMR study of single-molecule magnets reviewed above \cite{Harter_Chakov_Achey_Reyes_Kuhns_Christou_Dalal_2005} also exploited the ability of crystallites in polycrystalline samples to align in a magnetic field.

Finally, Song \etal\ \cite{Song_Kusumi_Kimura_Kimura_Deguchi_Ohki_Fujito_Simizu_2015} used a magnetically oriented microcrystal array of cellobiose to determine the magnitude and orientation of the $^{13}$C chemical shift tensors for selected carbons by MAS NMR of this sample.

\section{Conclusions and perspectives}

This review has presented an overview of instrumentation for SC NMR spectroscopy. While most studies rely on custom-made dedicated SC NMR probes, a few attempts have been made to circumvent this requirement by offering kits/modifications to adapt commercial probes for static solids or MAS for SC NMR experiments. Recent interesting studies focusing on orientation-specific excitation in MAS experiments may develop into promising alternatives for SC NMR experiments, although some work still needs to be done.

The theoretical basis for the analysis of SC NMR data has been briefly outlined and various pieces of software to analyse such experiments have been reviewed. It is reassuring that either dedicated SC NMR software or general-purpose solid-state NMR simulation software can be used to analyse SC NMR data implying that the barrier for analysing SC NMR data should not be too high.

Through an overview of a number of applications of SC NMR this review has demonstrated the versatility and relevance of SC NMR in various fields. Indeed, SC NMR is often not the first method of choice, since it is demanding to grow large crystals and record spectra for many orientations of the crystal. In spite of this, there is no doubt that SC NMR studies will be relevant also in the future, as SC NMR offers unique information about the orientation of the tensorial nuclear spin interactions that cannot be obtained otherwise.

%\section{Acknowledgements}

\printcredits

\bibliographystyle{model1-num-names}
%\bibliographystyle{model6-num-names}

\bibliography{MyLibrary}

\end{document}